\documentclass{article}

\usepackage{arxiv}

\usepackage[utf8]{inputenc} 
\usepackage[T1]{fontenc}    
\usepackage{hyperref}       
\usepackage{url}            
\usepackage{booktabs}       
\usepackage{amsfonts}       
\usepackage{nicefrac}       
\usepackage{microtype}      
\usepackage{multirow}
\usepackage{lipsum}
\usepackage{graphicx}
\usepackage{amsmath}
\usepackage{nccmath}
\usepackage[super]{nth}
\usepackage{float}
\graphicspath{ {./images/} }

\title{Unraveling the Temporal Importance of Community-scale Human Activity Features for Rapid Assessment of Flood Impacts}

\author{
 Faxi Yuan \\
  Urban Resilience.AI Lab\\
  Zachry Department of Civil and \\Environmental Engineering\\
  Texas A\&M University\\
  College Station, TX 77843 \\
  \texttt{faxi.yuan@tamu.edu} \\
  \And
 Yang Yang \\
  Urban Resilience.AI Lab\\
  Department of Computer Science and Engineering\\
  Texas A\&M University\\
  College Station, TX 77843 \\
  \texttt{yangyangsandy@tamu.edu} \\
  \And
 Qingchun Li \\
  Urban Resilience.AI Lab\\
  Zachry Department of Civil and \\Environmental Engineering\\
  Texas A\&M University\\
  College Station, TX 77843 \\
  \texttt{qingchunlea@tamu.edu} \\
  \And
 Ali Mostafavi \\
  Urban Resilience.AI Lab\\
  Zachry Department of Civil and \\Environmental Engineering\\
  Texas A\&M University\\
  College Station, TX 77843 \\
  \texttt{amostafavi@civil.tamu.edu} \\
}

\begin{document}
\maketitle
\begin{abstract}
The objective of this research is to explore the temporal importance of community-scale human activity features for rapid assessment of flood impacts. Ultimate flood impact data, such as flood inundation maps and insurance claims, becomes available only weeks and months after the floods have receded. Crisis response managers, however, need near-real-time data to prioritize emergency response. This time lag creates a need for rapid flood impact assessment. Accordingly, community-scale big data (such as satellite imagery) has been utilized for the early estimation of end flood impacts. Some recent studies have shown promising results for using human activity fluctuations as indicators of flood impacts. Existing studies, however, used mainly a single community-scale activity feature for the estimation of flood impacts and have not investigated their temporal importance for indicating flood impacts. Hence, in this study, we examined the importance of heterogeneous human activity features (such as human mobility, visits to points-of-interest, and social media posts) in different flood event stages. Using four community-scale big data categories we derived ten features related to the variations in human activity (e.g., travel, credit card transactions, and online communications) and evaluated their temporal importance for rapid assessment of flood impacts. Using multiple random forest models, we examined the temporal importance of each feature in indicating the extent of flood impacts (measured by the flood insurance claims and flood inundations) in the context of the 2017 Hurricane Harvey in Harris County, Texas. Our findings reveal that 1) fluctuations in human activity index and percentage of congested roads are the most important indicators for rapid flood impact assessment during response and recovery stages; 2) variations in credit card transactions assumed a middle ranking in both response and recovery stages; and 3) patterns of geolocated social media posts (Twitter) were of low importance across flood stages. Insights derived from data analysis reveal the potential for harnessing community-scale data characterizing human activity fluctuations for rapid assessment of flood impacts. The results of this research could rapidly forge a multi-tool enabling crisis managers to identify hotspots with severe flood impacts at various stages then to plan and prioritize effective response strategies.
\end{abstract}

\keywords{smart resilience \and urban flood \and machine learning \and big data \and rapid flood impact assessment}

\section*{Introduction}
As climate change induces more frequent extreme weather events, populated areas are at greater risk to destruction and disruption due to floods at many scales and at every facet of society. Floods cause significant social and physical impacts on communities, such as property loss, loss of life, damage to infrastructure, and disrupted access to critical facilities (Wang et al. 2020; Choi and Bae 2015; Huang et al. 2008; Downton and Pielke 2005). In the aftermath of a flood, rapid assessment of either inundated areas or damages could facilitate rapid identification of hotspots with more severe flood impacts (Kryvasheyeu et al. 2016) during flood event stages. Using hotspot information enables crisis response managers to prioritize recovery efforts and resource allocation to these areas (Ritter et al. 2020; Pappenberger et al. 2015). The ultimate flood impact data, such as flood inundation maps, insurance claims, and compiled household surveys, become available only several weeks and months after the flood events have ended (for example, Federal Emergency Management Agency flood inundation map for Hurricane Harvey in August 2017 became available in April 2018). This time lag makes the rapid assessment of flood impacts particularly critical.

Several studies have investigated the use of satellite imagery (Qiang et al. 2020; Skakun et al. 2014; Skakun 2010) and aerial images from drones (Akshya and Priyadarsini 2019; Baldazo et al. 2019; Popescu et al. 2015) for rapid flood impact assessment. Qiang et al. (2020) used nighttime satellite images to model flood impacts on and recovery of the economy. Skakun et al. (2014) analyzed the time series of satellite images to evaluate flood hazard and risk in Namibia. Popescu et al. (2015) used features such as color, texture and fractal types from drone-collected aerial images, to predict flood size, achieving accuracy of 98.87\%. While these data and assessments provide valuable insights regarding the extent and spatial variation of inundations, they have some limitations, such as their relatively coarse spatial and temporal resolution (Yuan and Liu 2018a) and higher computation costs for processing a large amount of imagery data (Morales-Alvarez et al. 2017). To enhance rapid flood impact assessment and complement the insights obtained from satellite and aerial images, researchers have investigated the usefulness of community-scale big data (Yabe et al. 2020a; Lu et al. 2016). The growth of sensing technologies and “data for good” programs of some technology companies have increased the availability of community-scale big data (Neelam and Sood 2020; Ianuale et al. 2015), such as activity data from cell phone signal densities, credit card transaction records, social media data and metadata, and traffic data. These community-scale big datasets have become more commonly available on the same day and at fine spatial resolution affording capture of human activities, such as daily activity indexes, transaction activities, online communications, and mobility. Large-scale flood-caused perturbations cause disruptions at the smaller scale of human activities (Podesta et al. 2021). Hence, variations in human activities in a flood-affected community signal impacts on the community (Farahmand et al. 2021), which can be further used for the rapid assessment of flood impacts (e.g., Fan et al. 2020; Yuan et al. 2021b). In particular, some recent studies have investigated single human activity features for rapid assessment of flood impacts. Fluctuations in the density of population activities (obtained through aggregate cell phone signals) or Waze road flooding reports could indicate local flood inundation status in an area outside a floodplain (Farahmand et al. 2021; Praharaj et al. 2021). Podesta et al. (2021) compared the fluctuations in visits to points of interest (POIs) during normal periods to the Hurricane Harvey period and found that such fluctuations can be used to assess flood impacts. Yuan et al. (2021b) assessed the flood impacts of Hurricane Harvey through the analysis of variations in credit card transactions. Kryvasheyeu et al. (2016) and Fan et al. (2020a) examined the Twitter posts of disaster-related tweets to estimate damage in Hurricane Sandy. Yuan and Liu (2020) evaluated both the posts related to a disaster and their sentiment expressions to examine damage by Hurricane Matthew. In addition, Fan et al. (2020b) have found that most roads with null values of average speed (from traffic data) in Harris County during Hurricane Harvey were actually inundated by the floods. This study provides promising evidence that changes in human mobility obtained from traffic data can also provide signals about flood impact. While these studies point to the promise of garnering insights from community-scale big data about heterogeneous human activity-based features, little is known about the temporal importance of these features across flood stages. For instance, changes in human mobility before an event might be due to preparedness and evacuation response of people and might not be a strong indication of flood impacts (Podesta et al. 2021). Changes in human mobility during and in the immediate aftermath could be a stronger signal about of flood disruption (Yabe et al. 2020a). Hence, it is essential to discern the temporal importance of human activity-based features for rapid assessment of flood impacts.

In pursuit of this goal, we examined ten features related to community’s activities, including daily activity index, transaction activities, online communications, and mobility in Harris County (Texas, USA) during the 2017 Hurricane Harvey. We investigated the fluctuations of these features from the normal period compared with the flood period and created a set of random forest models to explore the temporal changes in the importance of each feature across the flood stages. We present predicted results from one of the random forest models. It should be noted, however, that the purpose of this study was not to build a state-of-the-art prediction model of flood impacts based on these features, but rather to evaluate the temporal changes in the importance of features. 

\section*{Methods and Materials}

Figure 1 illustrates the research framework that guided the study. We first started by identifying aspects of community realities that are susceptible to external disruptions. Then we specified the community-scale data and associated features, which can capture the temporal variations in community realities. With the specified community-scale big data and associated features, we created multiple random forest models and utilized the function for feature importance analysis to analyze temporal changes in the importance of various features in terms of the extent to which they could indicate flood impacts. The ultimate flood impacts were based on the flood insurance claim data and flood inundation map across the 142 ZIP codes in Harris County.

\begin{figure}[ht]
\centering
\includegraphics[width=0.85\linewidth]{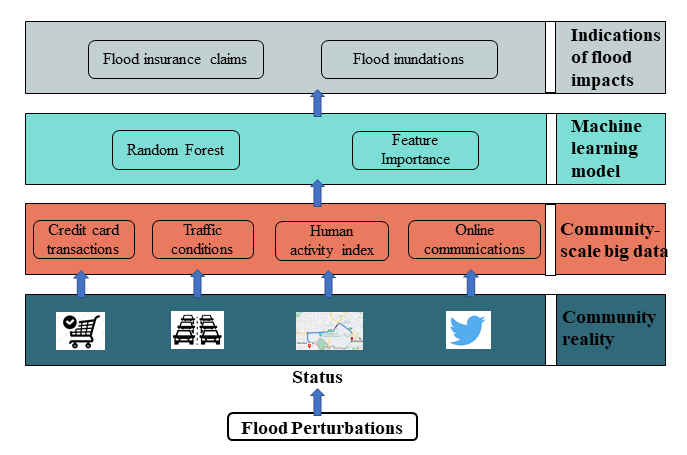}
\caption{Framework for examining temporal changes in the importance of human activity features for rapid assessment of flood impacts.}
\end{figure}

\subsection*{Community reality}
Community reality, the state of the day-to-day life of community (Curry 2016), can be captured based on various dimensions of human activities, such as mobility activities (Podesta et al. 2021; Yabe et al. 2020b), mobility and traffic patterns (Yabe at al. 2020c; Yuan et al. 2021a), credit card transactions (Yuan et al. 2021b), and online communications (Zhai et al. 2021). Human activities get perturbed due to direct effect on households, infrastructure damage, and people’s response behaviors. Hence, fluctuations in human activities could provide important insights regarding the state of community reality (Li et al. 2021). For example, human mobility activities were impacted by the COVID-19 pandemic (Yabe et al. 2020b). Podesta et al. (2021) showed that human visits to points of interests decreased during floods and thus the community had less mobility activities. In another study, Fan et al. (2020) showed that road inundation can be inferred from the absence of average speed data on road segments. Yuan et al. (2021b) examined the extent to which flood impacts were associated with the variations in credit card transactions. Online communications play an important role in collective sense-making of communities as disasters unfold. Evaluation of social media posts could be used for aggregating the personal toll (Zhang et al. 2020). When analyzing sentiment expressions on Twitter during Hurricane Harvey, Yuan et al. (2020) found that the impacted areas were more likely to indicate negative sentiment during disasters. Hence, the content of social media communications could shed insight on the dimension of human activities. (Fan et al. 2020c).

As a result, community reality and its variations (during crises compared with the normal period) can reveal signals of flood impacts. For instance, if residents are economically impacted by the floods, or if they could not access businesses such as restaurants and groceries due to inundated/closed roads, or if businesses such as pharmacies are closed due to damage, we can capture the flood impacts by analyzing community’s transaction behaviors (e.g., credit card transactions). In this study, we harnessed the community-scale big data related to different dimensions of community reality (i.e., human activity index, traffic and mobility, credit card transactions, and online social media communications) to explore the extent to which different features can indicate the flood impacts in the context of the 2017 Hurricane Harvey.

\subsection*{Community-scale big data}
\subsubsection*{\textit{Data description}}
To obtain features that signal flood impacts, we used four community-scale big data categories: (1) Mapbox data for human activity index, (2) IRINX traffic data, (3) SafeGraph credit card transaction data, and (4) Twitter data for collective sense making. For all datasets, the analysis timeline range is August 1, 2017, to September 15, 2017, as Hurricane Harvey made landfall in Harris County on August 25, 2017.

The Mapbox data contains indices of telemetry-based human activity that vary across space and time. The spatial unit of data aggregation is a tile. The partition of tiles is based on Mercantile, a Python library, which enables creating spatial-resolution grids. Human activity is collected, aggregated, and normalized by Mapbox based on the geography information updates of users’ cell phone locations. The more users located in a tile at time t, the greater the human activity index. The dataset includes human activity data for the entire United States. Mapbox provided the temporal resolution of 4 hours as the raw data. This dataset contains an activity metric for different tiles at different time points across the United States (${activity_{tile,t}}$), and the larger value of the activity metric ${activity_{tile,t}}$ reflects more human activities within that tile at time \emph{t}. Then, we aggregated the tiles (180 meter×180 meter) at the ZIP code level to derive the daily activity metric for our further analysis. Figure 2 shows examples for the activity index of ZIP codes before and after Hurricane Harvey approached Harris County. Polygons with colors ranging from light green to red represent the ZIP code with activity index changing from little to large values. When Hurricane Harvey made landfall on August 25, 2017 (Figure 2b), we can see activity index for more ZIP codes have increased as more orange, brown and red polygons appeared compared with that before the hurricane (i.e., August 16, 2017 in Figure 2a). In the response stage (e.g., August 30, 2017 I Figure 2c), more ZIP codes have seen the decrease of their activity index compared with Figure 2b. Compared with response stage in Figure 2c, we have seen more ZIP codes had less activity index in the recover stage (Figure 2d). Through Figures 2a-2d, we observed the variations of human activity index across flood stages, which could signal the flood impacts. 

\begin{figure}[ht]
\centering
\includegraphics[width=\linewidth]{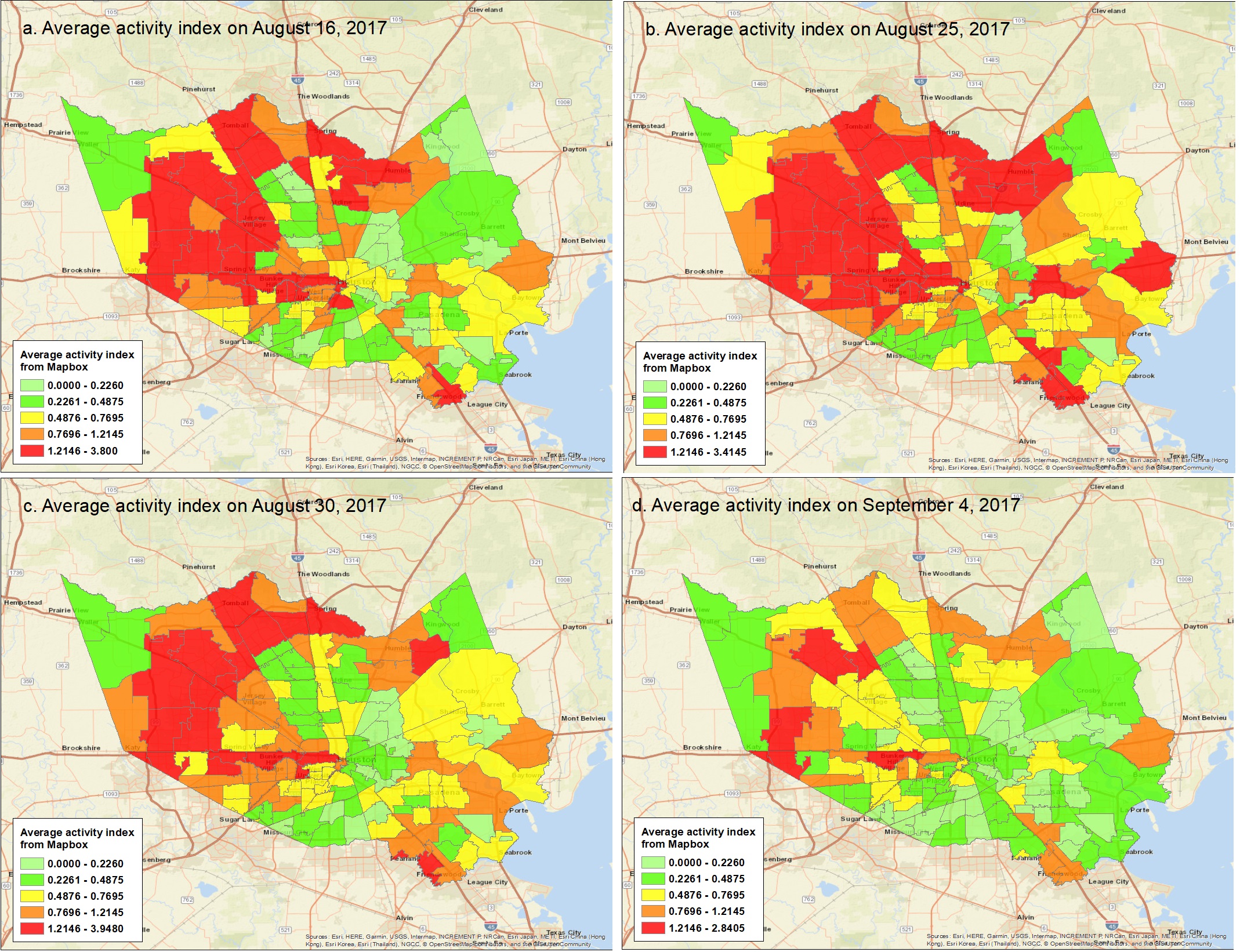}
\caption{Activity index of ZIP codes of Harris County on August 16, 2017 (2a), August 25, 2017 (2b), August 30, 2017 (2c), and September 4, 2017 (2d).}
\end{figure}

The INRIX dataset includes location-based traffic condition data from both sensors and vehicles at the road segment level. The INRIX traffic data contains the average traffic speed of each road segment at 5-minute intervals and their corresponding historical average traffic speed. The traffic data includes all available road segments—from interstates to intersections, and from country roads to neighborhoods. This dataset also provides road names, geographic locations defined its head and end coordinates, and length. We aggregated the road segments according to their locations by ZIP to extract features related to changes in traffic conditions.

Credit card transaction records data were obtained from SafeGraph. Each transaction record data contains the transaction date, cardholder’s residential ZIP code, the number of unique cards from the ZIP code involved in transactions on that day, the number of unique transactions on cards from the ZIP code observed on that day, and the total amount spent on cards from the ZIP code on that day. By matching the ZIP codes to transactions, we aggregated the daily credit card transaction activities.

The Twitter data was obtained from the Twitter API which enables streaming the geo-located tweets generated by its users from given periods and regions (Yuan and Liu 2018a). Each tweet contains username, tweet content, geolocations, and post time (Yuan et al. 2021c). Tweet geolocations by latitude and longitude were used to aggregate humans’ daily online communication activities by ZIP code.

\subsubsection*{\textit{Feature engineering}}
Using these four community-scale big data categories, we specified ten features (Table 1) to capture the community state during floods. This section describes how each feature was defined and calculated with given datasets. To examine signals of flood impacts, all features in Table 1 were evaluated based on their variations from the normal period (baseline period) to flood period. We defined the normal period from August 1, 2017, to August 24, 2017, for the features derived from Mapbox, IRINX, and SafeGraph datasets. For the Twitter dataset, we denoted the normal period from August 22, 2017, to August 24, 2017, due to the lack of Twitter data from August 1, 2017, to August 21, 2017.

\begin{table}[]
\caption{Summary of features derived from datasets}
\centering
\begin{tabular}{lll}
\hline
Dataset                    & Features                                                                                                           & Descriptions                                                                                                                                                                                                   \\ \hline
Mapbox                     & \begin{tabular}[c]{@{}l@{}}FE   1:  Variations in the \\ average daily   \\ activity index\end{tabular}            & \begin{tabular}[c]{@{}l@{}}The   percentage of variations of the average daily\\ activity of all tiles (180mx180m) within\\ each ZIP code compared with baseline.\end{tabular}                                 \\
\multirow{2}{*}{IRINX}     & \begin{tabular}[c]{@{}l@{}}FE 2:  Variations in the \\ daily maximum percentage \\ of congested roads\end{tabular} & \begin{tabular}[c]{@{}l@{}}The percentage of variations of the maximum percentage \\ of congested roads (speed below 50\% of speed limit) per \\ day within each ZIP code compared with baseline.\end{tabular} \\
                           & \begin{tabular}[c]{@{}l@{}}FE   3:  Changes in the \\ daily average percentage \\ of congested roads\end{tabular}  & \begin{tabular}[c]{@{}l@{}}The   percentage of variations of the mean percentage of \\ congested roads (speed below 50\% of speed limit) per day \\ within each ZIP code compared with baseline.\end{tabular}  \\
\multirow{3}{*}{SafeGraph} & \begin{tabular}[c]{@{}l@{}}FE 4:  Variations in the \\ number of cards\end{tabular}                                & \begin{tabular}[c]{@{}l@{}}The percentage of variations of the number of unique\\ credit cards used per day within each ZIP code compared \\ with baseline.\end{tabular}                                       \\
                           & \begin{tabular}[c]{@{}l@{}}FE 5:  Changes in the \\ number of transactions\end{tabular}                            & \begin{tabular}[c]{@{}l@{}}The percentage of variations of the number of \\ transactions per day within each ZIP code compared \\ with baseline.\end{tabular}                                                  \\
                           & \begin{tabular}[c]{@{}l@{}}FE   6:  Changes the total\\ spent\end{tabular}                                         & \begin{tabular}[c]{@{}l@{}}The   percentage of variations of the total spent per day\\ within each ZIP code   compared with baseline.\end{tabular}                                                             \\
\multirow{4}{*}{Twitter}   & \begin{tabular}[c]{@{}l@{}}FE 7:  Variations in the\\ average sentiment score\end{tabular}                         & \begin{tabular}[c]{@{}l@{}}The percentage of variations of the average sentiment \\ score of all geo-coordinated tweets per day within each \\ ZIP code compared with baseline.\end{tabular}                   \\
                           & \begin{tabular}[c]{@{}l@{}}FE 8:  Changes in the \\ number of positive tweets\end{tabular}                         & \begin{tabular}[c]{@{}l@{}}The percentage of variations of the number of tweets with\\ positive sentiment per day within each ZIP code compared \\ with baseline.\end{tabular}                                 \\
                           & \begin{tabular}[c]{@{}l@{}}FE 9:  Changes in the \\ number of neutral tweets\end{tabular}                          & \begin{tabular}[c]{@{}l@{}}The percentage of variations of the number of tweets with \\ neutral sentiment per day within each ZIP code compared \\ with baseline.\end{tabular}                                 \\
                           & \begin{tabular}[c]{@{}l@{}}FE 10:Changes in the \\ number of   negative tweets\end{tabular}                        & \begin{tabular}[c]{@{}l@{}}The percentage of variations of the number of tweets with\\ negative sentiment per day within each ZIP code compared \\ with baseline.\end{tabular}         \\ \hline                       
\end{tabular}
\end{table}

\textbf{Variations in the average daily activity index:} We computed the average daily activity index by averaging the daily activity indices for all tiles within each ZIP code. Accordingly, variations in the daily activity index values can be a potential indicator for the flood impacts (Farahmand et al, 2021). Using the mean of daily activity index in the normal period as the baseline, we calculated the variation of the average daily activity index during flood period (i.e., FE 1) using Equation (1), where \emph{t} represents the date.
\begin{equation}
\label{eq:1}
FE1=\frac{{mean\ of\ activity\ index}_{normal}-{activity\ index}_{t}}{{mean\ of\ activity\ index}_{normal}}\times100\%
\end{equation}

\textbf{Variations in the percentage of congested roads:} Using the IRINX traffic data, we specified two features: variations in the daily maximum percentage of congested roads (i.e., FE 2) and changes in the daily average percentage of congested roads (FE 3). According to the geographic location of each road segment, we divided road segments by ZIP codes. For each 5-minute period t within a day, we computed the ratio between the current average speed and the speed limit of that road segment (${ratio_{v_t,v_limit)}}$). In this study, we used 50\% speed loss to denote road congestion. If (${ratio_{v_t,v_limit)}}$) < 50\%, a road was denoted as congested. If an unflooded road is near a flooded road, this unflooded road is more likely to become congested (Fan et al. 2020a). Therefore, the variations in road congestion during flood periods can become a potential indicator for flood impacts. According to the status of road congestion, we computed the percentage of congested roads at period t within each ZIP code. Then, we calculated the maximum road congestion value and average road congestion value of the 288 percentages (1440 minutes per day/5 minutes per period) for each ZIP code. Using as baselines the normal maximum daily percentage of congested roads and the normal average daily percentage, we computed the fluctuations of daily maximum percentage of congested roads (i.e., FE 2) and the daily average percentage of congested roads (i.e., FE 3) during flood period with Equations (2) and (3), respectively, where \emph{t} represents the date.
\begin{equation}
\label{eq:2}
FE2=\frac{{maximum\ daily\ congestion\ percentage}_{normal}-{maximum\ daily\ congestion\ percentage}_{t}}{{maximum\ daily\ congestion\ percentage}_{normal}}\times100\%
\end{equation}

\begin{equation}
\label{eq:3}
FE3=\frac{{mean\ daily\ congestion\ percentage}_{normal}-{mean\ daily\ congestion\ percentage}_{t}}{{mean\ daily\ congestion\ percentage}_{normal}}\times100\%
\end{equation}

\textbf{Variations in the credit card transactions:} Using credit card transaction data, we calculated the number of cards, the number of transactions, and the total spent per day for each ZIP code. Yuan et al. (2021c) found that variations in credit card transactions from the normal period compared with the flood period can be an indicator of flood impacts. A significant negative fluctuation signals more severe flood impacts. The study by Yuan et al. (2021c) used the average daily spent in the normal period (i.e., three weeks before Hurricane Harvey made landfall) as the baseline from which to compute fluctuations of total spent during the flood period. Accordingly, we introduced two additional variables—the number of cards and the number of transactions—to calculate their variations as potential indicators for flood impacts. Specifically, using the averages of the daily number of cards, daily number of transactions, and daily total spent in the normal period, we computed the variations of these three variables for the daily values of features FE 4, FE 5, and FE 6 with Equations (4–6).
\begin{equation}
\label{eq:4}
FE4=\frac{{mean\ daily\ number\ of\ cards}_{normal}-{daily\ number\ of\ cards}_{t}}{{mean\ daily\ number\ of\ cards}_{normal}}\times100\%
\end{equation}

\begin{equation}
\label{eq:5}
FE5=\frac{{mean\ daily\ number\ of\ transactions}_{normal}-{daily\ number\ of\ transactions}_{t}}{{mean\ daily\ number\ of\ transactions}_{normal}}\times100\%
\end{equation}

\begin{equation}
\label{eq:6}
FE6=\frac{{mean\ daily\ total\ spent}_{normal}-{daily\ total\ spent}_{t}}{{mean\ daily\ total\ spent}_{normal}}\times100\%
\end{equation}

\textbf{Variations in the online communications on Twitter:} Existing studies have demonstrated the use of tweet sentiments for assessing disaster impacts in hurricanes and floods (Yuan and Liu 2020; Zou et al. 2018). Yuan and Liu (2020) employed variations in sentiment scores of Twitter posts from the normal period compared with the hurricane period to assess the disaster impacts and found an association between the sentiment score variations and flood impacts. In this study, we used the rule-based model called VADER (Hutto and Gilbert 2014) for the sentiment analysis with geolocated Twitter data. The VADER model calculates the normalized sentiment scores for Twitter data and provides a mechanism for denoting the sentiment polarities for a given text. If the normalized sentiment score $\geq$ 0.05, the given text data has positive sentiment; if the -0.05< normalized sentiment score <0.05, the given text data has neutral sentiment; and if normalized sentiment score $\leq$ -0.05, the given text data has negative sentiment. Using the VADER model, we computed the normalized sentiment score for each geolocated tweet. According to their geolocations, we aggregated the daily Twitter data for a ZIP code and calculated the daily average sentiment scores for all daily Twitter data within a ZIP code using Equation (7). Also, we determined the daily numbers of Twitter data with positive, neutral, and negative sentiment. With the average sentiment scores and numbers of tweets with positive, neutral, and negative sentiment in the normal period, we computed the daily values for FE 7, FE 8, FE 9, and FE 10 using equations (8-11). 
\begin{equation}
\label{eq:7}
average\ sentiment\ score=\frac{\sum_{i=1}^{{n}_{z,t}}{sentiment}_{t_i}}{{n}_{z,t}}
\end{equation}
where, $n_(z,t)$ represents the number of tweets within ZIP code \emph{z} on day \emph{t}; $sentiment_i$ denotes the normalized sentiment score of Twitter \emph{i}.

\begin{equation}
\label{eq:8}
FE7=\frac{{mean\ of\ daily\ average\ sentiment\ score}_{normal}-{daily\ average\ sentiment\ score}_{t}}{{mean\ of\ daily\ average\ sentiment\ score}_{normal}}\times100\%
\end{equation}

\begin{equation}
\label{eq:9}
FE8=\frac{{mean\ of\ daily\ number\ of\ positive\ tweets}_{normal}-{daily\ number\ of\ positive\ tweets}_{t}}{{mean\ of\ daily\ number\ of\ positive\ tweets}_{normal}}\times100\%
\end{equation}

\begin{equation}
\label{eq:10}
FE9=\frac{{mean\ of\ daily\ number\ of\ neutral\ tweets}_{normal}-{daily\ number\ of\ neutral\ tweets}_{t}}{{mean\ of\ daily\ number\ of\ neutral\ tweets}_{normal}}\times100\%
\end{equation}

\begin{equation}
\label{eq:11}
FE10=\frac{{mean\ of\ daily\ number\ of\ negative\ tweets}_{normal}-{daily\ number\ of\ negative\ tweets}_{t}}{{mean\ of\ daily\ number\ of\ negative\ tweets}_{normal}}\times100\%
\end{equation}

\subsection*{Random forest models}
Based on the concept of ensemble learning, random forest was developed as an extension of bagging technique with the trees-based algorithms. Compared with traditional machine learning approaches which learn one hypothesis from the training data, the ensemble method aggregates multiple hypotheses (Dietterich 2002) thereby reducing errors and variances within a single hypothesis. Given the concept of the ensemble method, the technique of bagging within the tree models can reduce the variances of a single decision tree (Oza and Russell 2011). The bagging technique divides the initial training dataset into several subsets then randomly substitutes subsets with replacements to train their corresponding decision trees. Thus, the bagging technique produces an ensemble of different tree models. Averaging all the prediction results from various tree models smooths out data and enables more reliable classifications.

Using the bagging technique, random forest modeling involves not only random selection of subset of training dataset but also the random selection of features within the training dataset (Prasad et al. 2006). Therefore, random forest can improve the variable selection based on the enhancement of bagging technique (Altman and Krzywinski 2017). In this study, we created daily random forest models to test the importance of our daily features (Table 1) for indicating flood impacts. We used the F1 score to evaluate the performance of these random forest models.

In addition, we utilized the feature importance evaluation method within the random forest modeling to assess the temporal changes in the importance of each feature. The random forest package from the \emph{scikit-learn} library uses the optimized version of the CART (classification and regression trees) algorithm to yield the largest information gain at each node measured by the Gini index (Daniya et al. 2020). Accordingly, we employed the aggregated decrease in Gini impurity to evaluate the feature importance for our daily random forest models. The aggregated decrease in Gini impurity can be calculated using Equations 12 through 16 (Kelly and Okada 2012). The descriptions of all the variables in these equations are summarized in Table 2.

\begin{equation}
\label{eq:12}
Gini(P_j)=\sum_{i=1}^{C}p_i(1-p_i)
\end{equation}

\begin{equation}
\label{eq:13}
\Delta Gini(j)=Gini(P_j )  -  \frac{|L_j |}{|P_j |} Gini(L_j ) -  \frac{|R_j |}{|P_j |} Gini(R_j) 
\end{equation}

\begin{equation}
\label{eq:14}
Imp(feature_i,t) = \sum _{j: node\ j\ splits\ by\ feature_i\  within\ a\ tree\ t}\Delta Gini(j)  
\end{equation}

\begin{equation}
\label{eq:15}
Norm\ Imp(feature_i,t) =  \frac{Imp(feature_i,t)}{\sum _{j\epsilon features\ within\ a\ tree\ t}Imp(feature_j)} 
\end{equation}

\begin{equation}
\label{eq:16}
RF\ Imp(feature_i) = \frac {\sum _{t\epsilon trees\ within\ a\ RF\ model}{Norm\ Imp(feature_i,t)}}{T} 
\end{equation}

\begin{table}[]
\caption{Summary of variables of equations 12 through 16}
\centering
\begin{tabular}{ll}
\hline
Variables              & Descriptions                                                                        \\ \hline
$P_j$                   & Partition   corresponding to node \emph(j) within a tree.                                  \\
$Gini(P_j)$            & Gini impurity of $P_j$.                                                              \\
\emph{C}                      & Number of classes.                                                                  \\
$p_i$                   & Probability that a randomly chosen case from $P_j$ will be a member of class \emph(i)       \\
$\delta Gini(j)$               & Reduction of Gini impurity at a non-terminal node \emph(j).                                \\
$L_j$ \& $R_j$           & Partitions corresponding to the left ($L_j$) and right ($R_j$) child nodes of node \emph{j}. \\
$Imp(feature_i,t)$      & Importance of $feature_i$ in a tree \emph{t}.                                               \\
$Norm Imp(feature_i,t)$ & Normalized importance of $feature_i$ in a tree \emph{t}.                                    \\
\emph{T}                      & Number of trees in a random forest model.                                           \\
$RF\ Imp(feature_i)$     & Importance of $feature_i$ in a random forest model.                          \\ \hline       
\end{tabular}
\end{table}

Using Equations 12 through 16, we determined the temporal variation in the features’ importance across daily random forest models. The daily changes in the features’ importance are examined for evaluating the time periods within which each feature to seek out signals regarding flood impacts.

\subsection*{Flood impacts }
Flood impacts as dependent variables of the random forest models are represented by two measures: (1) the normalized number of claims; (2) and the flood inundation percentages within a ZIP code. Specifically, with insurance claim data for Hurricane Harvey collected by National Flood Insurance Program (Mobley et al. 2021), we computed the number of claims for each ZIP code. Considering the effect of population size on the number of claims (Yuan and Liu 2018a), we used the normalized number of claims as one of our flood impact measures. The normalized number of claims was calculated based on the ratio between the number of claims and the population of that ZIP code from the US census data. Figure 3a shows results of the normalized number of claims. For the flood inundation percentages, we used the flood inundation map of Hurricane Harvey (Figure 3b) produced by Federal Emergency Management Administration. Overlapping this map with Harris County map at ZIP code level, we computed the flood inundation areas within a ZIP code and further calculated the flood inundation percentage for each ZIP code. Figure 3c shows the spatial distribution of the flood inundation percentages.

\begin{figure}[ht]
\centering
\includegraphics[width=\linewidth]{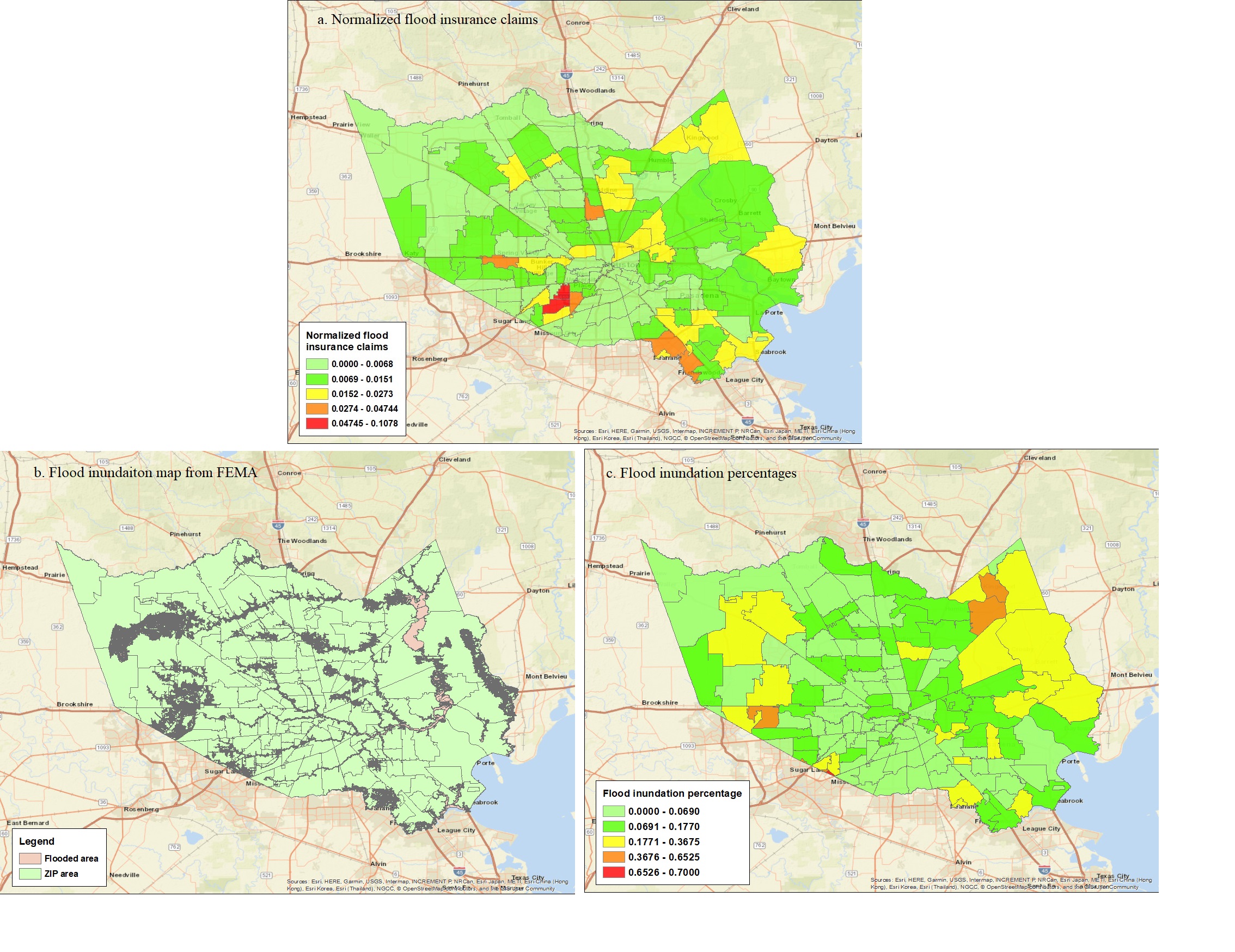}
\caption{Spatial distributions of normalized flood insurance claims (3a), flood inundation (3b), and flood inundation percentages (3c) by ZIP codes in the context of the 2017 Hurricane Harvey in Harris County. In these three figures, each polygon represents a ZIP code region.}
\end{figure}

With these two flood impact measures, we classified our ZIP codes with two to four classes of flood impacts, which are used as class labels for the ZIP codes. For the evaluation of temporal variations of features’ importance as indicators for flood impacts, we used these class labels as the input dependent variables for random forest feature importance function.

\section*{Results}
\subsection*{Feature importance for flood insurance claims}
This section summarizes the rank of feature importance for indicating the flood impacts (using flood insurance claims as a measure). With the normalized number of claims of the 142 ZIP codes, we classified the flood impacts into two, three or four classes. Taking three-class classification as an example, we performed the rank of the importance for 10 features (Figure 4). (See the rank of the importance for two-class and four-class classifications based on flood insurance claims in the Figures S1-S2 in the supplementary information). In these figures, orange curves represent human activities; green curves reflect the features derived from the credit card transactions; blue curves are for the travel activities derived from the traffic/road conditions, and pink curves represent online communications from the Twitter data. Given the dissipation date of Hurricane Harvey of September 2, 2017, we divided our study period (from August 25, 2017 to September 15, 2017), into two stages: 1) response stage: August 25, 2017–September 2, 2017 (9 days); and recovery stage: September 3, 2017–September 15, 2017 (13 days). This section takes the three-class classification of flood insurance claims as an example to discuss the rank of feature importance in the response and recovery stages. 

\begin{figure}[ht]
\centering
\includegraphics[width=\linewidth]{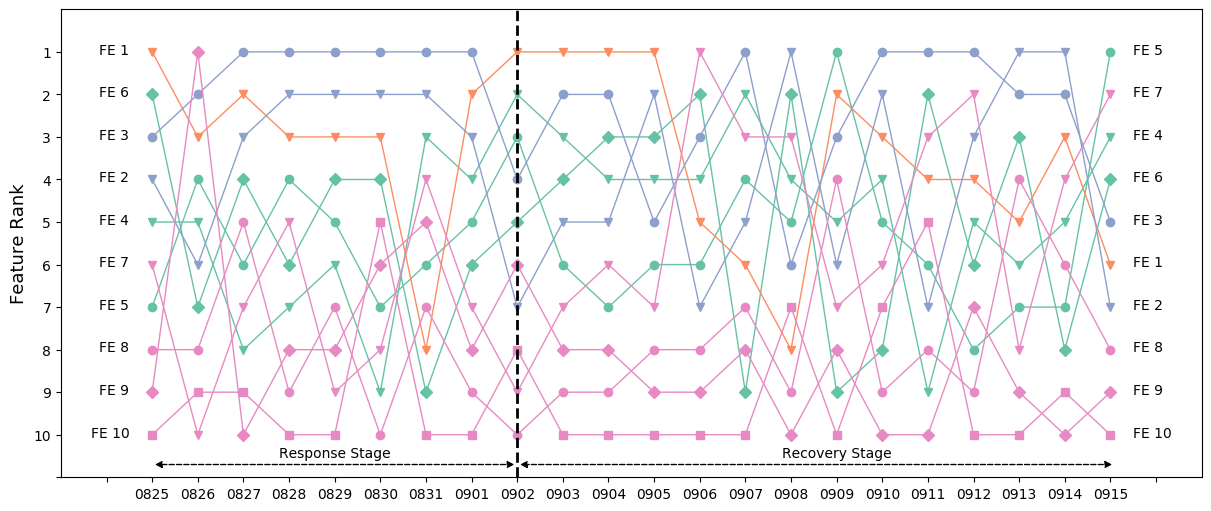}
\caption{Rank of feature importance for the three-class classification of flood insurance claims. In the training data, we used the 33\% and 66\% percentiles of the normalized number of claims of the 142 ZIP codes to classify claims into three classes for the flood impacts. If the normalized number of claims of a ZIP code is below the 33\% percentile, this ZIP code has slight flood impact; if it is between 34\% and 66\%, this ZIP code has medium flood impact; otherwise, this ZIP code has severe flood impact. The vertical black dashed line represents September 2, 2017, the date Hurricane Harvey dissipated. The left side of the vertical line represents the response stage; the right side stands for recovery stage. This information applies to Figures 5 and to S1–S4 (supplementary information).}
\end{figure}

\subsubsection*{\textit{Feature importance for indicating flood insurance claims in response stage}}
Table 3 shows the analysis results for the rank of feature importance for indicating flood impacts using insurance claims as the measure. For each feature as in Figure 4, we summarized the persistence period (the number of days the feature importance persisted and fluctuated slightly, second column, Table 3). The third column of Table 3 shows the rank persistence ranges for all the features in their corresponding persistence periods. The last column reveals the final rank of each feature calculated by the average of its ranks across the response stage. The feature importance analysis for the two-class and four-class classifications in the response stage are illustrated in Tables S1 and S2 in the supplementary information. 

\begin{table}[]
\caption{Analysis of feature importance for indicating three-class flood insurance claims in response stage (9 days)}
\centering
\begin{tabular}{llll}
\hline
Features                                                                                                            & \begin{tabular}[c]{@{}l@{}}Persistence \\ periods (days)\end{tabular} & \begin{tabular}[c]{@{}l@{}}Rank ranges in \\ persistence periods\end{tabular} & Final rank \\ \hline
\begin{tabular}[c]{@{}l@{}}FE 1:   Variations in the average \\ daily activity index\end{tabular}                   & 8                                                                     & 1–3                                                                           & 2          \\
\begin{tabular}[c]{@{}l@{}}FE 2:   Variations in the daily \\ maximum percentage of \\ congested roads\end{tabular} & 7                                                                     & 1–4                                                                           & 3          \\
\begin{tabular}[c]{@{}l@{}}FE 3:   Changes in the daily \\ average percentage of \\ congested roads\end{tabular}    & 8                                                                     & 1–3                                                                           & 1          \\
\begin{tabular}[c]{@{}l@{}}FE 4:   Variations in the number of \\ cards\end{tabular}                                & 7                                                                     & 3–7                                                                           & 4          \\
\begin{tabular}[c]{@{}l@{}}FE 5:   Changes in the number of \\ transactions\end{tabular}                            & 7                                                                     & 4–7                                                                           & 6          \\
FE 6:   Changes the total spent                                                                                     & 7                                                                     & 4–7                                                                           & 5          \\
\begin{tabular}[c]{@{}l@{}}FE 7:   Variations in the average \\ sentiment score\end{tabular}                        & 7                                                                     & 6–9                                                                           & 8          \\
\begin{tabular}[c]{@{}l@{}}FE 8:   Changes in the number of \\ positive tweets\end{tabular}                         & 7                                                                     & 7–10                                                                          & 9          \\
\begin{tabular}[c]{@{}l@{}}FE 9:   Changes in the number of \\ neutral tweets\end{tabular}                          & 7                                                                     & 5–9                                                                           & 7          \\
\begin{tabular}[c]{@{}l@{}}FE 10: Changes in the number of   \\ negative tweets\end{tabular}                        & 8                                                                     & 8–10                                                                          & 10         \\ \hline
\end{tabular}
\end{table}

According to Table 3, we can see humans’ daily activities (FE 1) and travel activities (FE 2 and FE 3) are more reliable indicators for flood impacts measured by flood insurance claims. During hurricane and flood periods, some of the affected residents may choose to stay at home, as they felt confident staying at home as they had ridden out previous extreme events (Bowser and Cutter 2015), while some could choose to follow the evacuation order (Dow and Cutter 2002). Either evacuating or sheltering in place can be captured by daily activity index and travel activities. For instance, staying at home could reduce travel activities, and evacuation could increase road congestion, which can result in the variations of average daily activity index and congested roads. Therefore, variations in the average daily activity index, daily maximum percentage of congested roads, and daily average percentage of congested roads, could provide indicators for rapid assessment of the extent of flood impacts in terms of insurance claims.

In addition, features related to credit card transactions (i.e., FE 4, FE 5, and FE 6) are among the middle rank from four to six. Yuan et al. (2021b) had found that variations in daily total expenditures from the normal period could capture the flood impact. The study showed that residents’ credit card transactions (total spent) decreased in business sectors such as drugstore, health care and groceries (Yuan et al. 2021b), which indicates that flood impacts can be captured by the changes in the credit card transactions. 

For the features derived from Twitter data (FE 7, FE 8, FE 9, and FE 10), changes in the number of neutral tweets (FE 9) demonstrated the greatest importance among the four features. Features’ relation to changes in the number of positive and negative sentiments show little importance as indicators for flood insurance claims. The quantity of Twitter data posted within a ZIP code region has a strong and positive relationship with the population of that region (Fan et al. 2020d). During Hurricane Harvey, less populated ZIP code regions posted limited Twitter posts, the division of which in three sentiment polarities could further reduce the number of Twitter data, which would result in null values for features FE 7, FE 8, FE 9, and FE 10. As a result, changes in the number of positive, negative, and neutral sentiments, as well as variations in the average sentiment score, may be less important indicators for flood impact assessment compared with other features derived from humans’ daily activities, travel activities and credit card transactions. 

\subsubsection*{\textit{Feature importance for indicating flood insurance claims in recovery stage}}
Table 4 shows results of analysis of the rank of feature importance (during the recovery stage) for indicating flood impacts in terms of insurance claims. The feature importance analysis for the two-class and four-class classifications based on flood insurances are illustrated in Tables S3-S4 in the supplementary information.

\begin{table}[]
\caption{Analysis of feature importance for indicating three-class flood insurance claims in recovery stage (13 days)}
\centering
\begin{tabular}{llll}
\hline
Features                                                                                                            & \begin{tabular}[c]{@{}l@{}}Persistence \\ periods (days)\end{tabular} & \begin{tabular}[c]{@{}l@{}}Rank ranges in \\ persistence periods\end{tabular} & Final rank \\ \hline
\begin{tabular}[c]{@{}l@{}}FE 1:   Variations in the average \\ daily activity index\end{tabular}                   & 9                                                                     & 1–5                                                                           & 2          \\
\begin{tabular}[c]{@{}l@{}}FE 2:   Variations in the daily \\ maximum percentage of \\ congested roads\end{tabular} & 10                                                                    & 1–5                                                                           & 2          \\
\begin{tabular}[c]{@{}l@{}}FE 3:   Changes in the daily \\ average percentage of \\ congested roads\end{tabular}    & 10                                                                    & 1–3                                                                           & 1          \\
\begin{tabular}[c]{@{}l@{}}FE 4:   Variations in the number of \\ cards\end{tabular}                                & 11                                                                    & 3–5                                                                           & 4          \\
\begin{tabular}[c]{@{}l@{}}FE 5:   Changes in the number of \\ transactions\end{tabular}                            & 10                                                                    & 4–7                                                                           & 5          \\
FE 6:   Changes the total spent                                                                                     & 10                                                                    & 2–8                                                                           & 7          \\
\begin{tabular}[c]{@{}l@{}}FE 7:   Variations in the average \\ sentiment score\end{tabular}                        & 12                                                                    & 1–7                                                                           & 6          \\
\begin{tabular}[c]{@{}l@{}}FE 8:   Changes in the number of \\ positive tweets\end{tabular}                         & 10                                                                    & 6–9                                                                           & 8          \\
\begin{tabular}[c]{@{}l@{}}FE 9:   Changes in the number of \\ neutral tweets\end{tabular}                          & 10                                                                    & 8–9                                                                           & 9          \\
\begin{tabular}[c]{@{}l@{}}FE 10: Changes in the number of   \\ negative tweets\end{tabular}                        & 10                                                                    & 9–10                                                                          & 10         \\ \hline
\end{tabular}
\end{table}

Compared with the rank of feature importance during the response stage for indicating flood insurance claims, there is no significant variation in the rank of features in the recovery stage. Features derived from humans’ daily activities (FE 1) and travel activities (FE 2 and FE 3) are still in the rank of top three places; features related to credit card transactions (i.e., FE 4, FE 5, and FE 6) are still at the middle ranks; and features related to Twitter activities are at the bottom of the rankings in terms of feature importance. This results indicate that the importance of features for rapid assessment of flood impacts (based on flood claims) does not vary significantly during response and recovery stages; the most important features during the response stage retain their importance ranking in the recovery stage. Hence, since these features are calculated each day, a daily value of each of the important features could be a reliable indicator of flood impacts.

\subsection*{Feature importance for flood inundations}
This section shows the rank of feature importance for indicating the flood impacts measured by the flood inundation. Based on the calculated flood inundation percentages of the 142 ZIP codes, we classified the flood impacts into two, three, and four classes. As in the section of \textbf{Feature importance for flood insurance claims}, we used the three-class classification as an example and conducted analysis of the rank of feature importance (Figure 5). 5 (See the rank of the importance for two-class and four-class classifications of flood inundations in Figures S3 and S4 in the supplementary information.) 

\begin{figure}[ht]
\centering
\includegraphics[width=\linewidth]{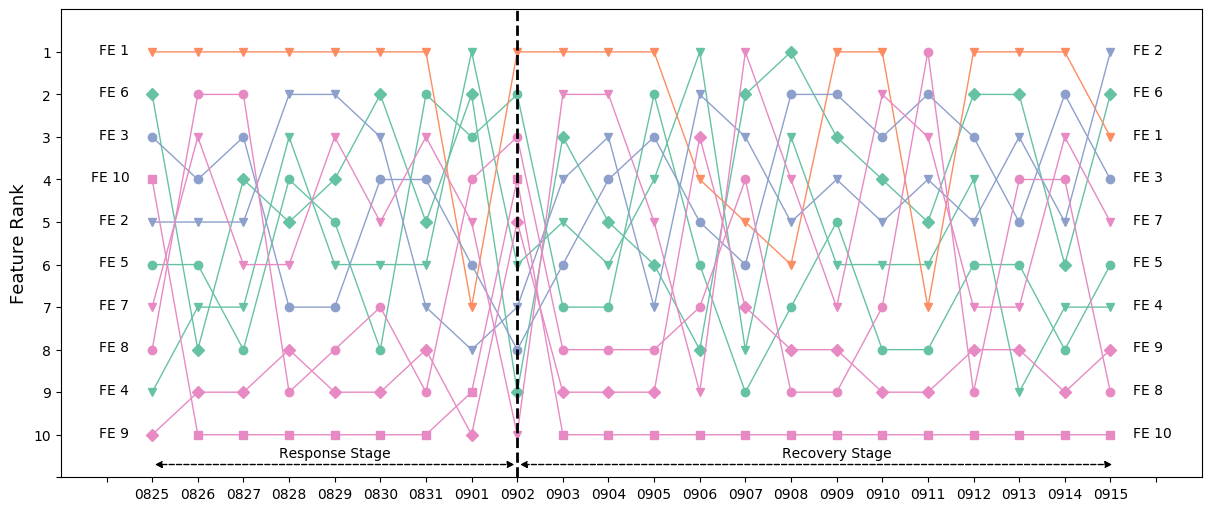}
\caption{Rank of feature importance for three-class classification of flood inundations. We defined ZIP codes with 0 flood percentage as the first class without flood impacts. Then, we used the 50\% percentile (i.e., median) of the flood percentages of the remaining ZIP codes (with flood percentage > 0) to classify them into another two classes for the flood impacts. If the flood percentage of a ZIP code is below the 50\% percentile, this ZIP code has slight (second class) flood impact; otherwise, this ZIP code has severe (third class) flood impact. In total, we had three classes of flood impacts: non-flood inundations (class 1), 0 < flood inundation percentage < median of flood inundation percentages (class 2), and median of flood inundation percentages $\leq $ flood inundation percentage (class 3).}
\end{figure}

\subsubsection*{Feature importance for indicating flood inundations in response stage}
Table 5 shows the analysis results for the rank of feature importance for predicting flood inundations in the response stage. For each feature presented in Figure 5, we summarized its persistence period, range of ranks in the persistence period, and its final rank. The feature importance analysis for the two-class and four-class classifications of flood inundations in the recovery stage are illustrated in Tables S5 and S6 in the supplementary information.

\begin{table}[]
\caption{Analysis of feature importance for indicating three-class flood inundations in response stage (9 days)}
\centering
\begin{tabular}{llll}
\hline
Features                                                                                                            & \begin{tabular}[c]{@{}l@{}}Persistence \\ periods (days)\end{tabular} & \begin{tabular}[c]{@{}l@{}}Rank ranges in \\ persistence periods\end{tabular} & Final rank \\ \hline
\begin{tabular}[c]{@{}l@{}}FE 1:   Variations in the average \\ daily activity index\end{tabular}                   & 8                                                                     & 1                                                                             & 1          \\
\begin{tabular}[c]{@{}l@{}}FE 2:   Variations in the daily \\ maximum percentage of \\ congested roads\end{tabular} & 8                                                                     & 2–7                                                                           & 3          \\
\begin{tabular}[c]{@{}l@{}}FE 3:   Changes in the daily \\ average percentage of \\ congested roads\end{tabular}    & 7                                                                     & 3–6                                                                           & 5          \\
\begin{tabular}[c]{@{}l@{}}FE 4:   Variations in the number of \\ cards\end{tabular}                                & 7                                                                     & 3–7                                                                           & 7          \\
\begin{tabular}[c]{@{}l@{}}FE 5:   Changes in the number of \\ transactions\end{tabular}                            & 7                                                                     & 2–6                                                                           & 4          \\
FE 6:   Changes the total spent                                                                                     & 7                                                                     & 2–5                                                                           & 2          \\
\begin{tabular}[c]{@{}l@{}}FE 7:   Variations in the average \\ sentiment score\end{tabular}                        & 7                                                                     & 3–6                                                                           & 6          \\
\begin{tabular}[c]{@{}l@{}}FE 8:   Changes in the number of \\ positive tweets\end{tabular}                         & 7                                                                     & 2–8                                                                           & 8          \\
\begin{tabular}[c]{@{}l@{}}FE 9:   Changes in the number of \\ neutral tweets\end{tabular}                          & 8                                                                     & 8–10                                                                          & 9          \\
\begin{tabular}[c]{@{}l@{}}FE 10: Changes in the number of   \\ negative tweets\end{tabular}                        & 8                                                                     & 9–10                                                                          & 10         \\ \hline
\end{tabular}
\end{table}

Compared with the rank of feature importance for indicating flood insurance claims in the response stage (Table 3), Table 5 shows some variations in the ranks when flood impacts are measured based on the flood inundations (such as variations of credit card transactions). In most cases, the general rank for feature importance rank remains stable when the measurement of flood impacts changes from flood insurance claims to flood inundations.  

\subsubsection*{Feature importance for indicating flood inundations claims in recovery stage}
Table 6 shows the analysis results for the rank of feature importance (during recovery stage) for indicating flood inundation extent. The feature importance analysis for the two-class and four-class classifications are illustrated in Tables S7–S8 in the supplementary information. 

\begin{table}[]
\caption{Analysis of feature importance for indicating three-class flood inundations in recovery stage (13 days)}
\centering
\begin{tabular}{llll}
\hline
Features                                                                                                            & \begin{tabular}[c]{@{}l@{}}Persistence \\ periods (days)\end{tabular} & \begin{tabular}[c]{@{}l@{}}Rank ranges in \\ persistence periods\end{tabular} & Final rank \\ \hline
\begin{tabular}[c]{@{}l@{}}FE 1:   Variations in the average \\ daily activity index\end{tabular}                   & 10                                                                    & 1–4                                                                           & 1          \\
\begin{tabular}[c]{@{}l@{}}FE 2:   Variations in the daily \\ maximum percentage of \\ congested roads\end{tabular} & 10                                                                    & 3–5                                                                           & 4          \\
\begin{tabular}[c]{@{}l@{}}FE 3:   Changes in the daily \\ average percentage of \\ congested roads\end{tabular}    & 11                                                                    & 2–5                                                                           & 2          \\
\begin{tabular}[c]{@{}l@{}}FE 4:   Variations in the number of \\ cards\end{tabular}                                & 10                                                                    & 4–7                                                                           & 6          \\
\begin{tabular}[c]{@{}l@{}}FE 5:   Changes in the number of \\ transactions\end{tabular}                            & 10                                                                    & 5–8                                                                           & 7          \\
FE 6:   Changes the total spent                                                                                     & 10                                                                    & 2–5                                                                           & 3          \\
\begin{tabular}[c]{@{}l@{}}FE 7:   Variations in the average \\ sentiment score\end{tabular}                        & 12                                                                    & 2–7                                                                           & 5          \\
\begin{tabular}[c]{@{}l@{}}FE 8:   Changes in the number of \\ positive tweets\end{tabular}                         & 12                                                                    & 4–9                                                                           & 8          \\
\begin{tabular}[c]{@{}l@{}}FE 9:   Changes in the number of \\ neutral tweets\end{tabular}                          & 12                                                                    & 7–9                                                                           & 9          \\
\begin{tabular}[c]{@{}l@{}}FE 10: Changes in the number of   \\ negative tweets\end{tabular}                        & 13                                                                    & 10                                                                            & 10         \\ \hline
\end{tabular}
\end{table}

Compared with the rank of feature importance indicating flood inundations in response stage (Table 5), Table 6 reveals minor variations in their rank during the recovery stage. The significant variations are captured in the rank changes in the daily average percentage of congested roads (FE 3) and changes of the number of transactions (FE 5). In the recovery stage, variation of the daily average percentage of congested roads becomes more important (rank changes from 5th place to 2nd place), while changes of the number of transactions become less important (rank changes from 4th place to 7th place). For the remaining eight features, variations of humans’ daily activities (FE 1) and travel activities (FE 2 and FE 3) are still more reliable indicators for flood inundations; changes of credit card transactions remain in the middle of the ranking (excluding FE 6 at 3rd place); and variations of online communications are still at bottom place (excluding FE 7 at 5th place).

Compared with importance rank of features indicating flood insurance claims (Table 4), we observe the main variations in the rank of the changes in credit card transactions feature when flood impacts are measured based on flood inundations (Table 6). The most significant variation falls in the importance rank of the changes of the total spent (FE 6), which moves from 7th to 3rd place. The importance rank of variations of the number of cards dropped from 4th place to 6th place and changes of the number of transactions dropped from 5th place to 7th place, when flood impacts reflected by insurance claims are changed to be measured by flood inundations. For the other seven features, there is no significant variation in their importance ranks. Hence, overall, the feature importance rankings for flood impact assessments related to two different flood impact measures—insurance claims and flood inundation extent—show very similar results. This result indicates that the features identified as being of top importance could provide reliable indicators of flood impacts when a rapid assessment is needed. 

\subsection*{Evaluations of random forest performance}
Given that the main goal of this study was to analyze the temporal importance of various features for indicating flood impacts, here we show an example for the evaluation of random forest performance. Taking the two-class classification of flood insurance claims as an example, we present the performance of the random forest model based on the F1 scores for days in (Figure 6). The F1 scores for the random forest models related to the three- and four-class classifications of flood insurance claims, as well as two-, three- and four-class classifications of flood inundations are illustrated in Figures S5 through S9 in the supplementary information. In Figure 6, the blue curve represents the F1 scores using the default parameters of the random forest model in the \emph{scikit-learn} library. Using the \emph{RandomsearchCV} package within the \emph{scikit-learn} library, we performed hyper parameter tuning to get the best performance of the random forest model. The results for the models with best parameters are illustrated in the yellow curve in Figure 6. We can observe the best performance (micro F1 score of 0.75) is for the model created based on the features of September 2, 2017 (dates of dissipation of Hurricane Harvey). 
\begin{figure}[ht]
\centering
\includegraphics[width=0.9\linewidth]{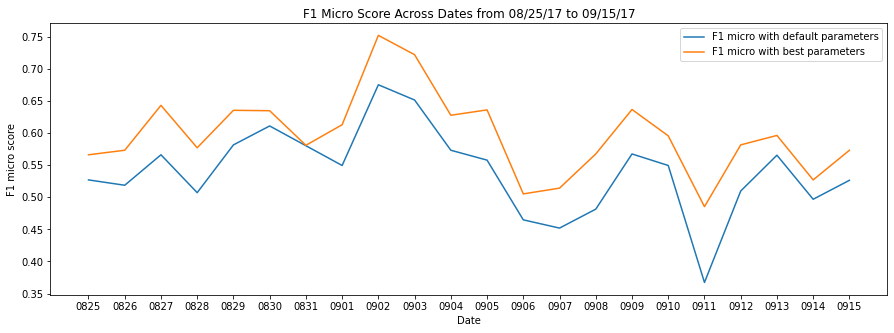}
\caption{Temporal F1 scores for random forest performances in the two-class classifications of flood insurance claims.}
\end{figure}

\section*{Discussions and Concluding Remarks}

Early and rapid estimation of flood inundations and losses across within a community can empower crisis response managers to identify areas with severe flood impacts to inform resource allocation during response and recovery. Flood inundation maps, insurance claims, and survey data reflecting flood impacts, however, become available only in weeks and months after the events. Emerging community-scale big data categories reveal fluctuations from the normal period to flood period, of community-scale activities, such as human activity index, travel activities, credit card transactions, and online communications, that could provide weak signals of flood impacts on the community. For instance, disruptions in infrastructure or population response behaviors could change the traffic and movements in the affected areas; thus, examining fluctuations in human activities and traffic could provide rapid and early indications of flood impacts.

While recent studies (e.g., Farahmand et al. 2021; Podesta et al. 2021; Yuan et al. 2021b; Zhang et al. 2020) demonstrated the potential of using human activity-based data for rapid impact assessment, the relative importance of features related to different aspects of human activities and their temporal significance was not known. Using four community-scale big data types, we derived ten features related to the changes in the daily human activity index, daily travel activities (reflected on road congestion conditions), daily credit card transactions, and daily online communications in the context of the 2017 Hurricane Harvey in Harris County. Through the use of the feature importance function within the random forest model, we explored the importance rank of these 10 features to indicate the ultimate flood impacts measured by both flood insurance claims and flood inundation across flood stages. With cases of three-class classifications of flood insurance claims and flood inundations, we found a stable rank of features derived from four categories of community-scale human activities. Features derived from the variations in the average daily activity index (FE 1), daily maximum percentage of congested roads (FE 2), and daily average percentage of congested roads (FE 3), are generally at the top three places in their importance rank in terms of indicating both flood insurance claims and flood inundations. Changes of credit card transactions in terms of the number of cards (FE 4), the number of transactions (FE 5), and the total spent (FE 6) generally in the middle of the importance scale in both response and recovery stages. Features derived from the variations of online communications on Twitter (FE 7, FE8, FE9, and FE10) mainly are the least relevant in terms of correlation with flood insurance claims and flood inundation in both stages among these four categories of community-scale human activities.

The study and findings contribute to the emerging field of smart resilience focusing on harnessing community-scale big data and analytics techniques to enhance disaster resilience capabilities, such as rapid impact assessment. Also, the findings could help public officials and emergency managers assess the impacts of floods before detained flood maps and claims become available. For instance, in both the response and recovery stages, crisis response managers could first use changes in the average daily activity index (FE 1) and the daily average percentage of congested roads (FE 3) as indicators of flood impacts, as these remain in the top-three ranked features when flood impacts are measured by flood insurance claims and flood inundations. If community-scale big data needed for determining features FE 1 and FE 3 are not available in the response stage, changes in the total spent (FE 6) could be another reliable indicator for flood impacts. With the identification of hotspots with severe flood impacts, crisis response managers can allocate relief resources to these areas. The analysis of feature importance rank could also suggest what features should be monitored by crisis response managers across different flood stages.

This research demonstrates that data-driven machine learning models with community-scale big data could provide important insights of flood impacts in discrete regions within a community and across flood stages through the fluctuations of community-scale human activities. One limitation in this study is the lack of consideration of relationships among these 10 features derived from four community-scale big data. We considered mainly the first-order feature importance and will investigate the dynamics of feature interactions (relationships) across flood stages in future research. The potential findings of their temporal relationships will inform which feature could impact another feature, which can help the selection of features to be monitored across flood stages.


\section*{Acknowledgements}
The authors would like to acknowledge funding support from the National Science Foundation CRISP 2.0 Type 2 \#1832662 and X-Grant project from the Texas A\&M University. The authors would also like to acknowledge INRIX, SafeGraph and Mapbox for providing the traffic data. Any opinions, findings, conclusions, or recommendations expressed in this research are those of the authors and do not necessarily reflect the view of the funding agencies.

\section*{Data availability}
The data that support the findings of this study are available from Mapbox, INRIX, SafeGraph, and Twitter, but restrictions apply to the availability of these data, which were used under license for the current study. The data can be accessed upon request submitted to each data provider. Other data (flood inundations and flood claims) used in this study are all publicly available.

\section*{Code availability}
The code that supports the findings of this study is available from the corresponding author upon request.

\section*{References}
Akshya, J., \& Priyadarsini, P. L. K. (2019). A hybrid machine learning approach for classifying aerial images of flood-hit areas. In 2019 International Conference on Computational Intelligence in Data Science (ICCIDS) (pp. 1-5). IEEE.

Altman, N., \& Krzywinski, M. (2017). Points of Significance: Ensemble methods: bagging and random forests. Nature Methods, 14(10), 933-935. 

Baldazo, D., Parras, J., \& Zazo, S. (2019). Decentralized multi-agent deep reinforcement learning in swarms of drones for flood monitoring. In 2019 27th European Signal Processing Conference (EUSIPCO) (pp. 1-5). IEEE.

Bowser, G. C., \& Cutter, S. L. (2015). Stay or go? Examining decision making and behavior in hurricane evacuations. Environment: Science and Policy for Sustainable Development, 57(6), 28-41.

Choi, S., \& Bae, B. (2015). The real-time monitoring system of social big data for disaster management. In Computer science and its applications (pp. 809-815). Springer, Berlin, Heidelberg.

Curry, K. (2016). Easily Confused Words: Realty vs. Reality. Available at < https://kathleenwcurry.wordpress.com>, last accessed on May 4, 2021.

Daniya, T., Geetha, M., \& Kumar, K. S. (2020). Classification And Regression Trees with Gini Index. Advances in Mathematics: Scientific Journal, 9(10), 8237-8247.

Dietterich, T. G. (2002). Ensemble learning. The handbook of brain theory and neural networks, 2, 110-125.

Dow, K., \& Cutter, S. L. (2002). Emerging hurricane evacuation issues: hurricane Floyd and South Carolina. Natural hazards review, 3(1), 12-18.

Downton, M. W., \& Pielke, R. A. (2005). How accurate are disaster loss data? The case of US flood damage. Natural Hazards, 35(2), 211-228.

Fan, C., Wu, F., \& Mostafavi, A. (2020a). A hybrid machine learning pipeline for automated mapping of events and locations from social media in disasters. IEEE Access, 8, 10478-10490.

Fan, C., Jiang, X., \& Mostafavi, A. (2020b). A network percolation-based contagion model of flood propagation and recession in urban road networks. Scientific Reports, 10(1), 1-12.

Fan, C., Shen, J., Mostafavi, A., \& Hu, X. (2020c). Characterizing reticulation in online social networks during disasters. Applied Network Science, 5(1), 1-20.

Fan, C., Esparza, M., Dargin, J., Wu, F., Oztekin, B., \& Mostafavi, A. (2020d). Spatial biases in crowdsourced data: Social media content attention concentrates on populous areas in disasters. Computers, Environment and Urban Systems, 83, 101514.

Farahmand, H., Wang, W., Maron, M., and Mostafavi, A. (2021). “Anomalous Human Activity Fluctuations from Digital Trace Data Signal Flood Inundation Status,” Arxiv.

Huang, X., Tan, H., Zhou, J., Yang, T., Benjamin, A., Wen, S. W., ... \& Li, X. (2008). Flood hazard in Hunan province of China: an economic loss analysis. Natural Hazards, 47(1), 65-73.

Hutto, C., \& Gilbert, E. (2014). Vader: A parsimonious rule-based model for sentiment analysis of social media text. In Proceedings of the International AAAI Conference on Web and Social Media (Vol. 8, No. 1).

Ianuale, N., Schiavon, D., \& Capobianco, E. (2015). Smart cities, big data, and communities: Reasoning from the viewpoint of attractors. IEEE Access, 4, 41-47.

Kelly, C., \& Okada, K. (2012). Variable interaction measures with random forest classifiers. In 2012 9th IEEE International Symposium on Biomedical Imaging (ISBI) (pp. 154-157). IEEE.

Kryvasheyeu, Y., Chen, H., Obradovich, N., Moro, E., Van Hentenryck, P., Fowler, J., \& Cebrian, M. (2016). Rapid assessment of disaster damage using social media activity. Science advances, 2(3), e1500779.

Li, Q., Tang, Z., Coleman, N., \& Mostafavi, A. (2021). Detecting Early-Warning Signals in Time Series of Visits to Points of Interest to Examine Population Response to COVID-19 Pandemic. IEEE Access, 9, 27189-27200.

Lu, X., Wrathall, D. J., Sundsøy, P. R., Nadiruzzaman, M., Wetter, E., Iqbal, A., ... \& Bengtsson, L. (2016). Unveiling hidden migration and mobility patterns in climate stressed regions: A longitudinal study of six million anonymous mobile phone users in Bangladesh. Global Environmental Change, 38, 1-7.

Mobley, W., Sebastian, A., Blessing, R., Highfield, W. E., Stearns, L., \& Brody, S. D. (2021). Quantification of continuous flood hazard using random forest classification and flood insurance claims at large spatial scales: a pilot study in southeast Texas. Natural Hazards and Earth System Sciences, 21(2), 807-822.

Neelam, S., \& Sood, S. K. (2020). A scientometric review of global research on smart disaster management. IEEE Transactions on Engineering Management, 68(1), 317-329.

Oza, N. C., \& Russell, S. J. (2001). Online bagging and boosting. In International Workshop on Artificial Intelligence and Statistics (pp. 229-236). PMLR.

Pappenberger, F., Cloke, H. L., Parker, D. J., Wetterhall, F., Richardson, D. S., \& Thielen, J. (2015). The monetary benefit of early flood warnings in Europe. Environmental Science \& Policy, 51, 278-291.

Podesta, C., Coleman, N., Esmalian, A., Yuan, F., \& Mostafavi, A. (2021). Quantifying community resilience based on fluctuations in visits to points-of-interest derived from digital trace data. Journal of the Royal Society Interface, 18(177), 20210158.

Popescu, D., Ichim, L., \& Caramihale, T. (2015). Flood areas detection based on UAV surveillance system. In 2015 19th International Conference on System Theory, Control and Computing (ICSTCC) (pp. 753-758). IEEE.

Praharaj, S., Chen, T. D., Zahura, F. T., Behl, M., \& Goodall, J. L. (2021). Estimating impacts of recurring flooding on roadway networks: a Norfolk, Virginia case study. Natural Hazards, 1-25.

Prasad, A. M., Iverson, L. R., \& Liaw, A. (2006). Newer classification and regression tree techniques: bagging and random forests for ecological prediction. Ecosystems, 9(2), 181-199.

Qiang, Y., Huang, Q., \& Xu, J. (2020). Observing community resilience from space: Using nighttime lights to model economic disturbance and recovery pattern in natural disaster. Sustainable Cities and Society, 57, 102115.

Ritter, J., Berenguer, M., Corral, C., Park, S., \& Sempere-Torres, D. (2020). ReAFFIRM: Real-time assessment of flash flood impacts–a regional high-resolution method. Environment international, 136, 105375.

Skakun, S., Kussul, N., Shelestov, A., \& Kussul, O. (2014). Flood hazard and flood risk assessment using a time series of satellite images: A case study in Namibia. Risk Analysis, 34(8), 1521-1537.

Skakun, S. (2010). A neural network approach to flood mapping using satellite imagery. Computing and Informatics, 29(6), 1013-1024.

Wang, Y., Li, J., Zhao, X., Feng, G., \& Luo, X. R. (2020). Using Mobile Phone Data for Emergency Management: a Systematic Literature Review. Information Systems Frontiers, 1-21.

Yabe, T., Zhang, Y., \& Ukkusuri, S. V. (2020a). Quantifying the economic impact of disasters on businesses using human mobility data: a Bayesian causal inference approach. EPJ Data Science, 9(1), 36.

Yabe, T., Tsubouchi, K., Fujiwara, N., Wada, T., Sekimoto, Y., \& Ukkusuri, S. V. (2020b). Non-compulsory measures sufficiently reduced human mobility in Tokyo during the COVID-19 epidemic. Scientific reports, 10(1), 1-9.

Yabe, T., Tsubouchi, K., Fujiwara, N., Sekimoto, Y., \& Ukkusuri, S. V. (2020c). Understanding post-disaster population recovery patterns. Journal of the Royal Society Interface, 17(163), 20190532.

Yuan, F., Xu, Y., Li, Q., \& Mostafavi, A. (2021a). Spatio-Temporal Graph Convolutional Networks for Road Network Inundation Status Prediction during Urban Flooding. arXiv preprint arXiv:2104.02276.

Yuan, F., Esmalian, A., Oztekin, B., \& Mostafavi, A. (2021b). Unveiling Spatial Patterns of Disaster Impacts and Recovery Using Credit Card Transaction Variances. arXiv preprint arXiv:2101.10090.

Yuan, F., Li, M., Liu, R., Zhai, W., \& Qi, B. (2021c). Social media for enhanced understanding of disaster resilience during Hurricane Florence. International Journal of Information Management, 57, 102289.

Yuan, F., Li, M., \& Liu, R. (2020). Understanding the evolutions of public responses using social media: Hurricane Matthew case study. International Journal of Disaster Risk Reduction, 51, 101798.

Yuan, F., \& Liu, R. (2020). Mining social media data for rapid damage assessment during Hurricane Matthew: Feasibility study. Journal of Computing in Civil Engineering, 34(3), 05020001.

Yuan, F., \& Liu, R. (2018a). Feasibility study of using crowdsourcing to identify critical affected areas for rapid damage assessment: Hurricane Matthew case study. International journal of disaster risk reduction, 28, 758-767.

Yuan, F., \& Liu, R. (2018b). Crowdsourcing for forensic disaster investigations: Hurricane Harvey case study. Natural Hazards, 93(3), 1529-1546.

Zhai, W., Peng, Z. R., \& Yuan, F. (2020). Examine the effects of neighborhood equity on disaster situational awareness: Harness machine learning and geotagged Twitter data. International Journal of Disaster Risk Reduction, 48, 101611.

Zhang, C., Yao, W., Yang, Y., Huang, R., \& Mostafavi, A. (2020). Semiautomated social media analytics for sensing societal impacts due to community disruptions during disasters. Computer‐Aided Civil and Infrastructure Engineering.

Zou, L., Lam, N. S., Cai, H., \& Qiang, Y. (2018). Mining Twitter data for improved understanding of disaster resilience. Annals of the American Association of Geographers, 108(5), 1422-1441.

\section*{Supplemental Information}
\appendix
\counterwithin{figure}{section}

\begin{table}[h]
\renewcommand\thetable{S1} 
   \centering
   \caption{Analysis of feature importance for indicating two-class flood insurance claims in response period (9 days)}
\begin{tabular}{llll}
\hline
Features                                                                                                            & \begin{tabular}[c]{@{}l@{}}Persistence \\ periods (days)\end{tabular} & \begin{tabular}[c]{@{}l@{}}Rank ranges in \\ persistence periods\end{tabular} & Final rank \\ \hline
\begin{tabular}[c]{@{}l@{}}FE 1:   Variations in the average \\ daily activity index\end{tabular}                   & 8                                                                     & 1–3                                                                           & 1          \\
\begin{tabular}[c]{@{}l@{}}FE 2:   Variations in the daily \\ maximum percentage of \\ congested roads\end{tabular} & 8                                                                     & 2–4                                                                           & 3          \\
\begin{tabular}[c]{@{}l@{}}FE 3:   Changes in the daily \\ average percentage of \\ congested roads\end{tabular}    & 8                                                                     & 1–2                                                                           & 2          \\
\begin{tabular}[c]{@{}l@{}}FE 4:   Variations in the number of \\ cards\end{tabular}                                & 8                                                                     & 3–6                                                                           & 4          \\
\begin{tabular}[c]{@{}l@{}}FE 5:   Changes in the number of \\ transactions\end{tabular}                            & 7                                                                     & 4–8                                                                           & 5          \\
FE 6:   Changes the total spent                                                                                     & 7                                                                     & 5–7                                                                           & 6          \\
\begin{tabular}[c]{@{}l@{}}FE 7:   Variations in the average \\ sentiment score\end{tabular}                        & 7                                                                     & 6–9                                                                           & 9          \\
\begin{tabular}[c]{@{}l@{}}FE 8:   Changes in the number of \\ positive tweets\end{tabular}                         & 7                                                                     & 7–9                                                                           & 7          \\
\begin{tabular}[c]{@{}l@{}}FE 9:   Changes in the number of \\ neutral tweets\end{tabular}                          & 8                                                                     & 6–9                                                                           & 7          \\
\begin{tabular}[c]{@{}l@{}}FE 10: Changes in the number of   \\ negative tweets\end{tabular}                        & 10                                                                    & 9–10                                                                          & 10         \\ \hline
\end{tabular}
\end{table}

\begin{table}[]
\renewcommand\thetable{S2} 
   \centering
   \caption{Analysis of feature importance for indicating four-class flood insurance claims in response period (9 days)}
\begin{tabular}{llll}
\hline
Features                                                                                                            & \begin{tabular}[c]{@{}l@{}}Persistence \\ periods (days)\end{tabular} & \begin{tabular}[c]{@{}l@{}}Rank ranges in \\ persistence periods\end{tabular} & Final rank \\ \hline
\begin{tabular}[c]{@{}l@{}}FE 1:   Variations in the average \\ daily activity index\end{tabular}                   & 8                                                                     & 1–2                                                                           & 1          \\
\begin{tabular}[c]{@{}l@{}}FE 2:   Variations in the daily \\ maximum percentage of \\ congested roads\end{tabular} & 7                                                                     & 3–7                                                                           & 4          \\
\begin{tabular}[c]{@{}l@{}}FE 3:   Changes in the daily \\ average percentage of \\ congested roads\end{tabular}    & 8                                                                     & 1–3                                                                           & 2          \\
\begin{tabular}[c]{@{}l@{}}FE 4:   Variations in the number of \\ cards\end{tabular}                                & 8                                                                     & 4–6                                                                           & 6          \\
\begin{tabular}[c]{@{}l@{}}FE 5:   Changes in the number of \\ transactions\end{tabular}                            & 9                                                                     & 3–7                                                                           & 5          \\
FE 6:   Changes the total spent                                                                                     & 7                                                                     & 2–5                                                                           & 3          \\
\begin{tabular}[c]{@{}l@{}}FE 7:   Variations in the average \\ sentiment score\end{tabular}                        & 7                                                                     & 5–9                                                                           & 7          \\
\begin{tabular}[c]{@{}l@{}}FE 8:   Changes in the number of \\ positive tweets\end{tabular}                         & 8                                                                     & 7–9                                                                           & 9          \\
\begin{tabular}[c]{@{}l@{}}FE 9:   Changes in the number of \\ neutral tweets\end{tabular}                          & 8                                                                     & 6–9                                                                           & 8          \\
\begin{tabular}[c]{@{}l@{}}FE 10: Changes in the number of   \\ negative tweets\end{tabular}                        & 9                                                                     & 10                                                                            & 10         \\ \hline
\end{tabular}
\end{table}

\begin{table}[]
\renewcommand\thetable{S3} 
   \centering
   \caption{Analysis of feature importance for indicating two-class flood insurance claims in recovery stage (13 days)}
\begin{tabular}{llll}
\hline
Features                                                                                                            & \begin{tabular}[c]{@{}l@{}}Persistence \\ periods (days)\end{tabular} & \begin{tabular}[c]{@{}l@{}}Rank ranges in \\ persistence periods\end{tabular} & Final rank \\ \hline
\begin{tabular}[c]{@{}l@{}}FE 1:   Variations in the average \\ daily activity index\end{tabular}                   & 12                                                                    & 1–6                                                                           & 3          \\
\begin{tabular}[c]{@{}l@{}}FE 2:   Variations in the daily \\ maximum percentage of \\ congested roads\end{tabular} & 10                                                                    & 1–4                                                                           & 2          \\
\begin{tabular}[c]{@{}l@{}}FE 3:   Changes in the daily \\ average percentage of \\ congested roads\end{tabular}    & 11                                                                    & 1–4                                                                           & 1          \\
\begin{tabular}[c]{@{}l@{}}FE 4:   Variations in the number of \\ cards\end{tabular}                                & 10                                                                    & 3–6                                                                           & 5          \\
\begin{tabular}[c]{@{}l@{}}FE 5:   Changes in the number of \\ transactions\end{tabular}                            & 10                                                                    & 2–6                                                                           & 4          \\
FE 6:   Changes the total spent                                                                                     & 11                                                                    & 4–8                                                                           & 6          \\
\begin{tabular}[c]{@{}l@{}}FE 7:   Variations in the average \\ sentiment score\end{tabular}                        & 13                                                                    & 5–9                                                                           & 7          \\
\begin{tabular}[c]{@{}l@{}}FE 8:   Changes in the number of \\ positive tweets\end{tabular}                         & 10                                                                    & 6–9                                                                           & 8          \\
\begin{tabular}[c]{@{}l@{}}FE 9:   Changes in the number of \\ neutral tweets\end{tabular}                          & 10                                                                    & 8–10                                                                          & 8          \\
\begin{tabular}[c]{@{}l@{}}FE 10: Changes in the number of   \\ negative tweets\end{tabular}                        & 13                                                                    & 9-10                                                                          & 10         \\ \hline
\end{tabular}
\end{table}

\begin{table}[]
\renewcommand\thetable{S4} 
   \centering
   \caption{Analysis of feature importance for indicating four-class flood insurance claims in recovery stage (13 days)}
\begin{tabular}{llll}
\hline
Features                                                                                                            & \begin{tabular}[c]{@{}l@{}}Persistence \\ periods (days)\end{tabular} & \begin{tabular}[c]{@{}l@{}}Rank ranges in \\ persistence periods\end{tabular} & Final rank \\ \hline
\begin{tabular}[c]{@{}l@{}}FE 1:   Variations in the average \\ daily activity index\end{tabular}                   & 10                                                                    & 1–3                                                                           & 1          \\
\begin{tabular}[c]{@{}l@{}}FE 2:   Variations in the daily \\ maximum percentage of \\ congested roads\end{tabular} & 10                                                                    & 1–5                                                                           & 3          \\
\begin{tabular}[c]{@{}l@{}}FE 3:   Changes in the daily \\ average percentage of \\ congested roads\end{tabular}    & 10                                                                    & 1–3                                                                           & 1          \\
\begin{tabular}[c]{@{}l@{}}FE 4:   Variations in the number of \\ cards\end{tabular}                                & 12                                                                    & 3–8                                                                           & 7          \\
\begin{tabular}[c]{@{}l@{}}FE 5:   Changes in the number of \\ transactions\end{tabular}                            & 11                                                                    & 3–7                                                                           & 5          \\
FE 6:   Changes the total spent                                                                                     & 12                                                                    & 3–9                                                                           & 4          \\
\begin{tabular}[c]{@{}l@{}}FE 7:   Variations in the average \\ sentiment score\end{tabular}                        & 11                                                                    & 4–7                                                                           & 5          \\
\begin{tabular}[c]{@{}l@{}}FE 8:   Changes in the number of \\ positive tweets\end{tabular}                         & 11                                                                    & 8–9                                                                           & 9          \\
\begin{tabular}[c]{@{}l@{}}FE 9:   Changes in the number of \\ neutral tweets\end{tabular}                          & 12                                                                    & 5–9                                                                           & 8          \\
\begin{tabular}[c]{@{}l@{}}FE 10: Changes in the number of   \\ negative tweets\end{tabular}                        & 12                                                                    & 10                                                                            & 10         \\ \hline
\end{tabular}
\end{table}

\begin{table}[]
\renewcommand\thetable{S5} 
   \centering
   \caption{Analysis of feature importance for indicating two-class flood inundations in response stage (9 days)}
\begin{tabular}{llll}
\hline
Features                                                                                                            & \begin{tabular}[c]{@{}l@{}}Persistence \\ periods (days)\end{tabular} & \begin{tabular}[c]{@{}l@{}}Rank ranges in \\ persistence periods\end{tabular} & Final rank \\ \hline
\begin{tabular}[c]{@{}l@{}}FE 1:   Variations in the average \\ daily activity index\end{tabular}                   & 8                                                                     & 1                                                                             & 1          \\
\begin{tabular}[c]{@{}l@{}}FE 2:   Variations in the daily \\ maximum percentage of \\ congested roads\end{tabular} & 8                                                                     & 3–8                                                                           & 4          \\
\begin{tabular}[c]{@{}l@{}}FE 3:   Changes in the daily \\ average percentage of \\ congested roads\end{tabular}    & 8                                                                     & 2–6                                                                           & 3          \\
\begin{tabular}[c]{@{}l@{}}FE 4:   Variations in the number of \\ cards\end{tabular}                                & 9                                                                     & 4–7                                                                           & 5          \\
\begin{tabular}[c]{@{}l@{}}FE 5:   Changes in the number of \\ transactions\end{tabular}                            & 7                                                                     & 5–8                                                                           & 7          \\
FE 6:   Changes the total spent                                                                                     & 8                                                                     & 2–4                                                                           & 2          \\
\begin{tabular}[c]{@{}l@{}}FE 7:   Variations in the average \\ sentiment score\end{tabular}                        & 7                                                                     & 4–9                                                                           & 6          \\
\begin{tabular}[c]{@{}l@{}}FE 8:   Changes in the number of \\ positive tweets\end{tabular}                         & 7                                                                     & 5–8                                                                           & 8          \\
\begin{tabular}[c]{@{}l@{}}FE 9:   Changes in the number of \\ neutral tweets\end{tabular}                          & 7                                                                     & 9                                                                             & 9          \\
\begin{tabular}[c]{@{}l@{}}FE 10: Changes in the number of   \\ negative tweets\end{tabular}                        & 8                                                                     & 10                                                                            & 10         \\ \hline
\end{tabular}
\end{table}

\begin{table}[]
\renewcommand\thetable{S6} 
   \centering
   \caption{Analysis of feature importance for indicating four-class flood inundations in response stage (9 days)}
\begin{tabular}{llll}
\hline
Features                                                                                                            & \begin{tabular}[c]{@{}l@{}}Persistence \\ periods (days)\end{tabular} & \begin{tabular}[c]{@{}l@{}}Rank ranges in \\ persistence periods\end{tabular} & Final rank \\ \hline
\begin{tabular}[c]{@{}l@{}}FE 1:   Variations in the average \\ daily activity index\end{tabular}                   & 7                                                                     & 1–2                                                                           & 1          \\
\begin{tabular}[c]{@{}l@{}}FE 2:   Variations in the daily \\ maximum percentage of \\ congested roads\end{tabular} & 7                                                                     & 2–6                                                                           & 4          \\
\begin{tabular}[c]{@{}l@{}}FE 3:   Changes in the daily \\ average percentage of \\ congested roads\end{tabular}    & 8                                                                     & 2–4                                                                           & 2          \\
\begin{tabular}[c]{@{}l@{}}FE 4:   Variations in the number of \\ cards\end{tabular}                                & 9                                                                     & 3–8                                                                           & 6          \\
\begin{tabular}[c]{@{}l@{}}FE 5:   Changes in the number of \\ transactions\end{tabular}                            & 7                                                                     & 2–7                                                                           & 5          \\
FE 6:   Changes the total spent                                                                                     & 7                                                                     & 1–5                                                                           & 3          \\
\begin{tabular}[c]{@{}l@{}}FE 7:   Variations in the average \\ sentiment score\end{tabular}                        & 7                                                                     & 6–9                                                                           & 8          \\
\begin{tabular}[c]{@{}l@{}}FE 8:   Changes in the number of \\ positive tweets\end{tabular}                         & 7                                                                     & 3–7                                                                           & 7          \\
\begin{tabular}[c]{@{}l@{}}FE 9:   Changes in the number of \\ neutral tweets\end{tabular}                          & 9                                                                     & 7–10                                                                          & 9          \\
\begin{tabular}[c]{@{}l@{}}FE 10: Changes in the number of   \\ negative tweets\end{tabular}                        & 8                                                                     & 10                                                                            & 10         \\ \hline
\end{tabular}
\end{table}

\begin{table}[]
\renewcommand\thetable{S7} 
   \centering
   \caption{Analysis of feature importance for indicating two-class flood inundations in recovery stage (13 days)}
\begin{tabular}{llll}
\hline
Features                                                                                                            & \begin{tabular}[c]{@{}l@{}}Persistence \\ periods (days)\end{tabular} & \begin{tabular}[c]{@{}l@{}}Rank ranges in \\ persistence periods\end{tabular} & Final rank \\ \hline
\begin{tabular}[c]{@{}l@{}}FE 1:   Variations in the average \\ daily activity index\end{tabular}                   & 12                                                                    & 1–6                                                                           & 3          \\
\begin{tabular}[c]{@{}l@{}}FE 2:   Variations in the daily \\ maximum percentage of \\ congested roads\end{tabular} & 12                                                                    & 1–6                                                                           & 2          \\
\begin{tabular}[c]{@{}l@{}}FE 3:   Changes in the daily \\ average percentage of \\ congested roads\end{tabular}    & 12                                                                    & 1–4                                                                           & 1          \\
\begin{tabular}[c]{@{}l@{}}FE 4:   Variations in the number of \\ cards\end{tabular}                                & 11                                                                    & 3–9                                                                           & 6          \\
\begin{tabular}[c]{@{}l@{}}FE 5:   Changes in the number of \\ transactions\end{tabular}                            & 13                                                                    & 5–8                                                                           & 9          \\
FE 6:   Changes the total spent                                                                                     & 11                                                                    & 2–5                                                                           & 4          \\
\begin{tabular}[c]{@{}l@{}}FE 7:   Variations in the average \\ sentiment score\end{tabular}                        & 11                                                                    & 2–8                                                                           & 5          \\
\begin{tabular}[c]{@{}l@{}}FE 8:   Changes in the number of \\ positive tweets\end{tabular}                         & 10                                                                    & 7–9                                                                           & 8          \\
\begin{tabular}[c]{@{}l@{}}FE 9:   Changes in the number of \\ neutral tweets\end{tabular}                          & 11                                                                    & 5–9                                                                           & 7          \\
\begin{tabular}[c]{@{}l@{}}FE 10: Changes in the number of   \\ negative tweets\end{tabular}                        & 13                                                                    & 10                                                                            & 10         \\ \hline
\end{tabular}
\end{table}

\begin{table}[]
\renewcommand\thetable{S8} 
   \centering
   \caption{Analysis of feature importance for indicating four-class flood inundations in recovery stage (13 days)}
\begin{tabular}{llll}
\hline
Features                                                                                                            & \begin{tabular}[c]{@{}l@{}}Persistence \\ periods (days)\end{tabular} & \begin{tabular}[c]{@{}l@{}}Rank ranges in \\ persistence periods\end{tabular} & Final rank \\ \hline
\begin{tabular}[c]{@{}l@{}}FE 1:   Variations in the average \\ daily activity index\end{tabular}                   & 10                                                                    & 1–6                                                                           & 4          \\
\begin{tabular}[c]{@{}l@{}}FE 2:   Variations in the daily \\ maximum percentage of \\ congested roads\end{tabular} & 10                                                                    & 1–4                                                                           & 1          \\
\begin{tabular}[c]{@{}l@{}}FE 3:   Changes in the daily \\ average percentage of \\ congested roads\end{tabular}    & 12                                                                    & 2–6                                                                           & 2          \\
\begin{tabular}[c]{@{}l@{}}FE 4:   Variations in the number of \\ cards\end{tabular}                                & 10                                                                    & 2–7                                                                           & 5          \\
\begin{tabular}[c]{@{}l@{}}FE 5:   Changes in the number of \\ transactions\end{tabular}                            & 12                                                                    & 4–9                                                                           & 8          \\
FE 6:   Changes the total spent                                                                                     & 10                                                                    & 1–5                                                                           & 2          \\
\begin{tabular}[c]{@{}l@{}}FE 7:   Variations in the average \\ sentiment score\end{tabular}                        & 10                                                                    & 2–8                                                                           & 6          \\
\begin{tabular}[c]{@{}l@{}}FE 8:   Changes in the number of \\ positive tweets\end{tabular}                         & 10                                                                    & 4–8                                                                           & 6          \\
\begin{tabular}[c]{@{}l@{}}FE 9:   Changes in the number of \\ neutral tweets\end{tabular}                          & 13                                                                    & 4–10                                                                          & 9          \\
\begin{tabular}[c]{@{}l@{}}FE 10: Changes in the number of   \\ negative tweets\end{tabular}                        & 13                                                                    & 9–10                                                                          & 10         \\ \hline
\end{tabular}
\end{table}

\begin{figure}[ht]
\renewcommand\thefigure{S1} 
\centering
\includegraphics[width=\linewidth]{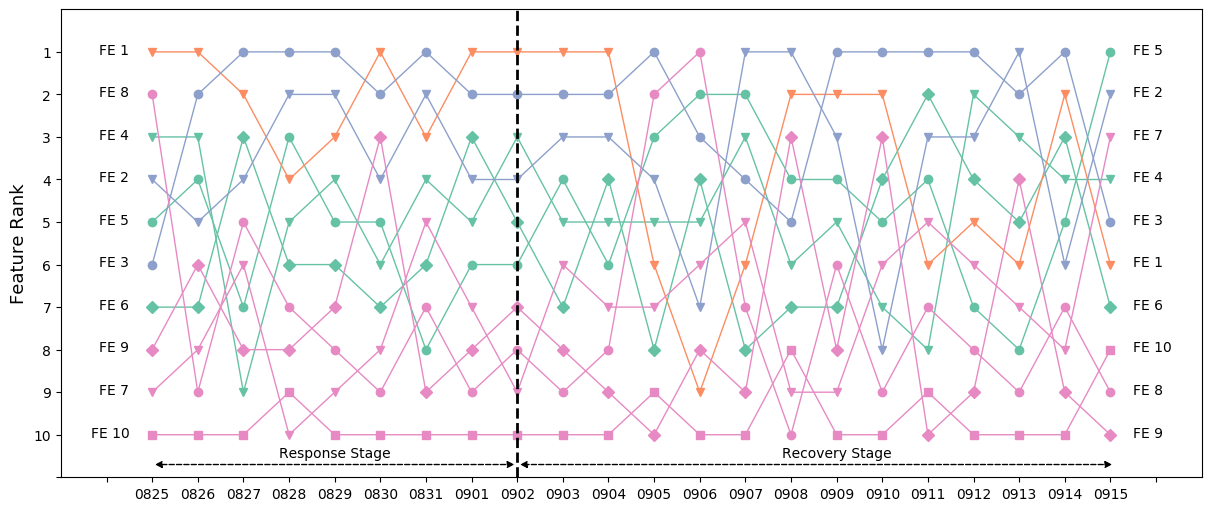}
\caption{Rank of feature importance for two-class classification of flood insurance claims. In the training data, we used the 50\% percentile (i.e., median) of the normalized number of claims of the 142 ZIP codes to classify them into two classes for the flood impacts. If the normalized number of claims of a ZIP code is below the 50\% percentile, this ZIP code has slight flood impact; otherwise, this ZIP code has severe flood impact.}
\end{figure}

\begin{figure}[ht]
\renewcommand\thefigure{S2} 
\centering
\includegraphics[width=\linewidth]{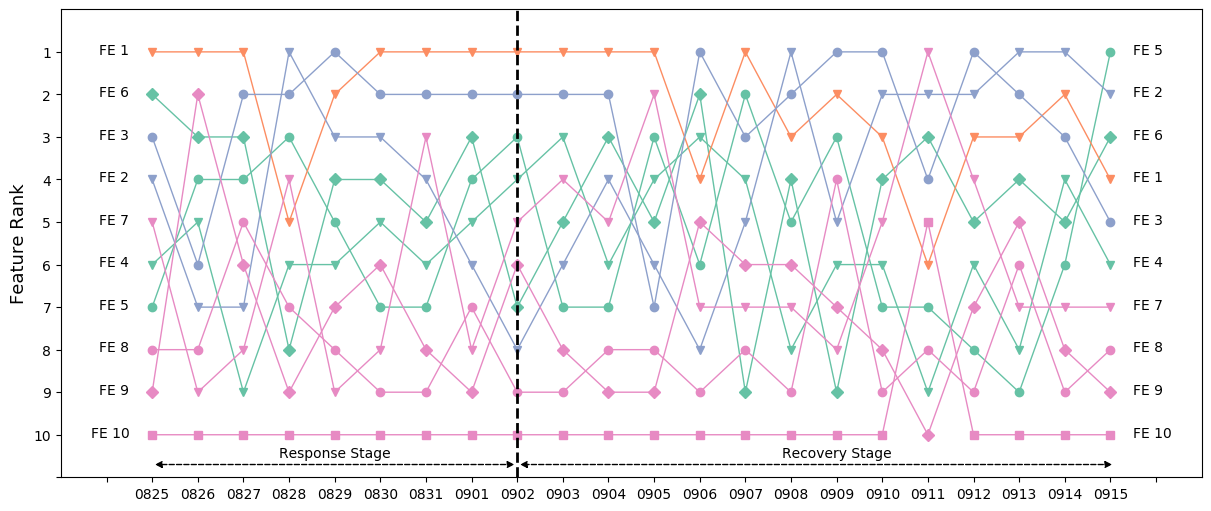}
\caption{Rank of feature importance for four-class classification of flood insurance claims. In the training data, we used the 25\%, 50\%, and 75\% percentiles of the normalized number of claims of the 142 ZIP codes to classify them into four classes for the flood impacts. If the normalized number of claims of a ZIP code is below the 25\% percentile, this ZIP code has very slight flood impact; if that is between 26\% and 50\%, this ZIP code has slight flood impact; if that is between 51\% and 75\%, this ZIP code has medium flood impact; otherwise, this ZIP code has severe flood impact.}
\end{figure}

\begin{figure}[ht]
\renewcommand\thefigure{S3} 
\centering
\includegraphics[width=\linewidth]{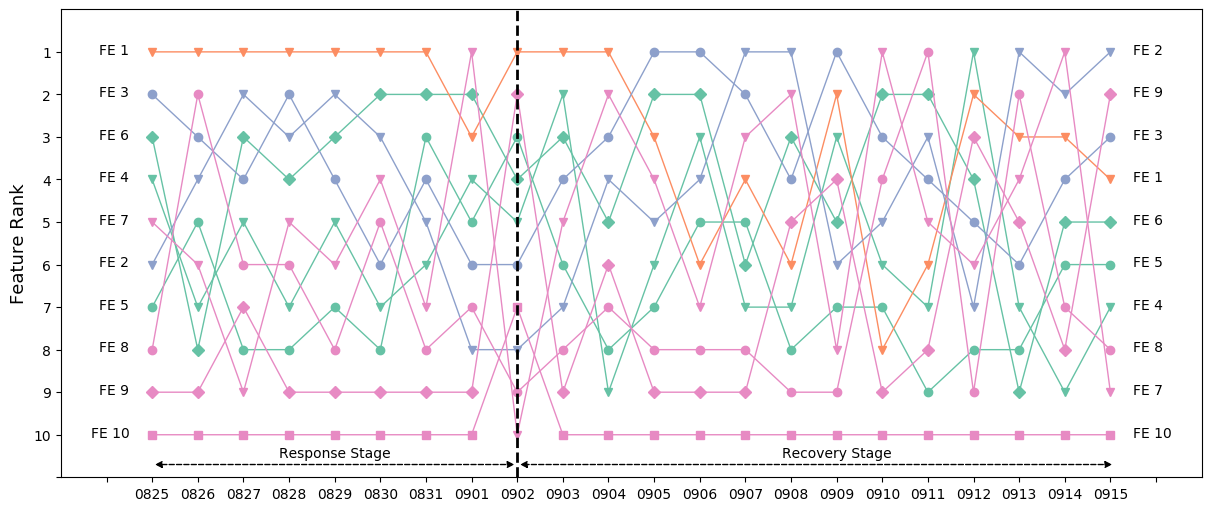}
\caption{Rank of feature importance for two-class classification of flood inundations. Using the flood inundation percentage for each ZIP code, we classified the 142 ZIP codes into two classes. If the flood inundation percentage is larger than 0, this ZIP code has flood impact; otherwise, this ZIP code does not have flood impact.}
\end{figure}

\begin{figure}[ht]
\renewcommand\thefigure{S4} 
\centering
\includegraphics[width=\linewidth]{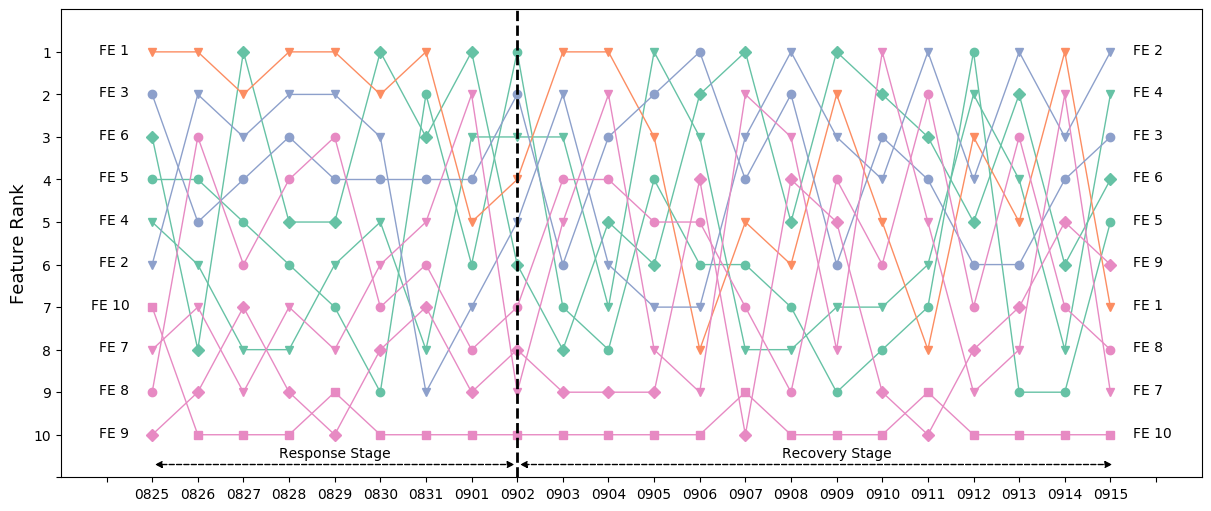}   
\caption{Rank of feature importance for four-class classification of flood inundations. In addition to the ZIP codes with 0 flood percentage (i.e., first class without flood impacts), we used the 33\% and 66\% percentiles of the flood percentages of the remaining ZIP codes to classify them into another three classes for the flood impacts. If the flood percentage of a ZIP code is below the 33\% percentile, this ZIP code has slight flood impact (i.e., second class); if that is between 34\% and 66\%, this ZIP code has medium flood impact (i.e., third class); otherwise, this ZIP code has severe flood impact (i.e., fourth class).}
\end{figure}

\begin{figure}[ht]
\renewcommand\thefigure{S5} 
\centering
\includegraphics[width=\linewidth]{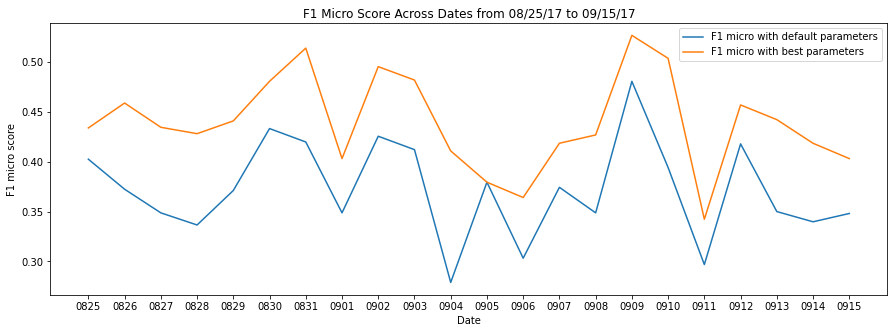}
\caption{Temporal F1 scores for random forest performances in the three-class classifications of flood insurance claims.}
\end{figure}

\begin{figure}[ht]
\renewcommand\thefigure{S6} 
\centering
\includegraphics[width=\linewidth]{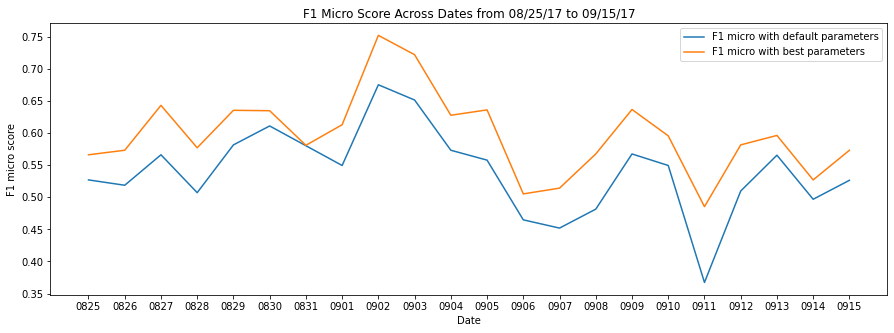}
\caption{Temporal F1 scores for random forest performances in the four-class classifications of flood insurance claims.}
\end{figure}

\begin{figure}[ht]
\renewcommand\thefigure{S7} 
\centering
\includegraphics[width=\linewidth]{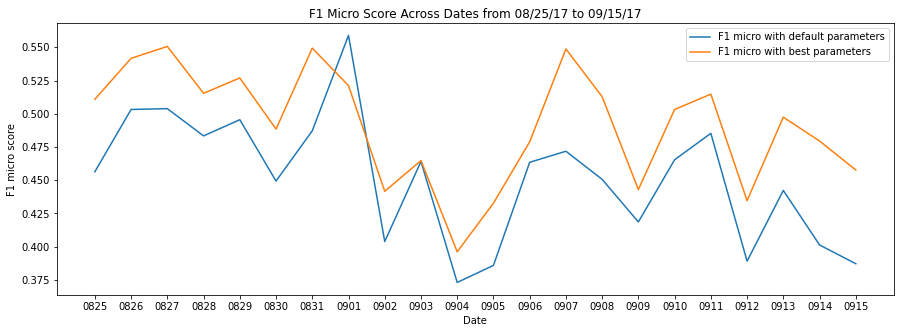}   
\caption{Temporal F1 scores for random forest performances in the two-class classifications of flood inundations.}
\end{figure}

\begin{figure}[ht]
\renewcommand\thefigure{S8} 
\centering
\includegraphics[width=\linewidth]{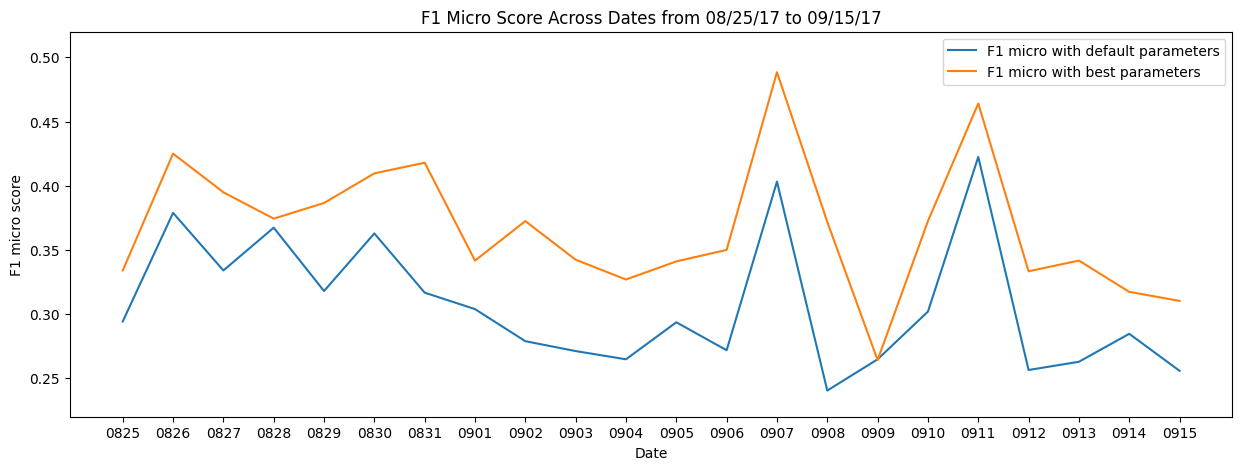}
\caption{Temporal F1 scores for random forest performances in the three-class classifications of flood inundations.}
\end{figure}

\begin{figure}[ht]
\renewcommand\thefigure{S9} 
\centering
\includegraphics[width=\linewidth]{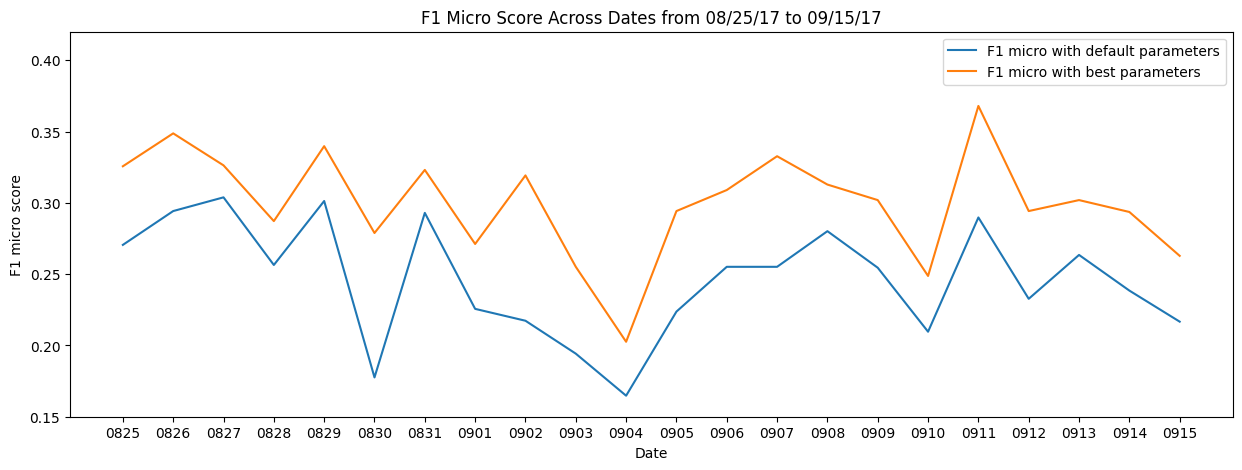}
\caption{Temporal F1 scores for random forest performances in the four-class classifications of flood inundations.}
\end{figure}

\end{document}